\documentclass[12pt]{article}
\usepackage{epsfig}
\usepackage{epstopdf}
\usepackage{amsmath}
\usepackage{amsfonts}
\usepackage{amssymb}
\input amssym.def
\input amssym.tex

\def\be{\begin{equation}}
\def\ee{\end{equation}}

\def\bea{\begin{eqnarray}}
\def\eea{\end{eqnarray}}

\def\ba{\begin{array}}
\def\ea{\end{array}}
\def\bd{\begin{displaymath}}
\def\ed{\end{displaymath}}

\def\Tr{{\rm Tr}}

\def\unit{1 \hskip-.3em \raise2pt\hbox{$ \scriptstyle |$ } }

\def\a{\alpha}
\def\b{\beta}

\def\d{\delta}
\def\e{\epsilon}           
  
\def\g{\gamma}

\def\l{\lambda}
\def\m{\mu}
\def\n{\nu}
\def\o{\omega}  
  \def\th{\theta}                  
\def\r{\rho}                                     
\def\s{\sigma}                                   
\def\t{\tau}

\def\D{\Delta}

\def\G{\Gamma}

\def\O{\Omega}

\def\cc{{\cal C}}

\def\cl{{\cal L}}

\def\cn{{\cal N}}
\def\co{{\cal O}}

\def\bd{\begin{displaymath}}
\def\ed{\end{displaymath}}
\def\6{\partial}
\def\N4{{\cal N}=4}
\def\lab{\label}

\newcommand{\UU}{a}
\newcommand{\Uprime}{a'}
\newcommand{\eps}{\epsilon}
\newcommand{\ga}{\gamma}
\newcommand{\Lag}{\mathcal{L}}
\newcommand{\rom}[1]{\mathrm{#1}}

\def\half{{1 \over 2}}

\def\bop#1{\setbox0=\hbox{$#1M$}\mkern1.5mu
        \vbox{\hrule height0pt depth.04\ht0
        \hbox{\vrule width.04\ht0 height.9\ht0 \kern.9\ht0
        \vrule width.04\ht0}\hrule height.04\ht0}\mkern1.5mu}
\def\pa{\partial}                              

\def\>{\rangle} 

\def\<{\langle} 
\def\Dsl{D \hskip-.6em \raise1pt\hbox{$ / $ } }
\def\to{\rightarrow}

\def\pa{\partial}

\def\+{\oplus}

\newcommand{\reef}[1]{(\ref{#1})}

\def\half{{1 \over 2}}
\def\Tr{{\rm Tr}\, }

\begin{document}
\begin{titlepage}

\begin{flushright}
MIT-CTP-3823 \\
{\tt hep-th/0703201}
\end{flushright}
\vspace{0.5cm}

\begin{center}
{\Large\bf From Fake Supergravity to Superstars}
\vspace{1cm}

{\bf Henriette Elvang${}^{b}$,
Daniel Z. Freedman${}^{a\,b}$
 and Hong Liu${}^{b}$} \\
\vspace{0.7cm}
 { ${}^{a}${\it Department of Mathematics}\\
${}^{b}${\it Center for Theoretical Physics}\\
 {\it Massachusetts Institute of Technology}\\
{\it Cambridge MA 02139, USA} }\\[5mm]
{\small \tt elvang@lns.mit.edu, dzf@math.mit.edu, hong\_liu@mit.edu}

\end{center}
\vskip .3truecm

\begin{abstract}
The fake supergravity method is applied to 5-dimensional
asymptotically AdS spacetimes containing gravity coupled to a real
scalar and an abelian gauge field.
The motivation is to obtain bulk solutions with 
$\mathbb{R}\times S^3$ symmetry
in order to explore the AdS/CFT correspondence
when the boundary gauge theory is on $\mathbb{R}\times S^3$. A fake
supergravity action, invariant under local supersymmetry through
linear order in fermion fields, is obtained. The gauge field makes
things more restrictive than in previous applications of fake
supergravity which allowed quite general scalar potentials. Here the
superpotential must take the form $W(\phi) \sim \exp(-k\phi) +
c~\exp(\frac{2\phi}{3k})$, and the only freedom is the choice
of the constant $k$. The fermion
transformation rules of fake supergravity lead to fake
Killing spinor equations. From their
integrability conditions, we obtain first order differential equations
which we solve analytically to find singular electrically charged 
solutions of the Lagrangian field equations. A Schwarzschild mass term
can be added to produce a horizon which shields the singularity. 
The solutions, which include ``superstars'', turn out to be known in
the literature. We compute their holographic parameters.
\end{abstract}

\end{titlepage}

\tableofcontents

\setcounter{equation}{0}
\section{Introduction}
The AdS/CFT correspondence has stimulated the study of
asymptotically anti-de Sitter spacetimes in various dimensions.
Quite often these spacetimes are solutions of a supergravity theory
containing gravity coupled to bosonic matter fields. In this
setting, it is common to search first for BPS solutions which
support Killing spinors. The BPS conditions are first order
differential equations which are frequently easier to solve than the
Lagrangian field equations. BPS solutions  have residual
supersymmetry. They are a small subset of the solutions one would
like to study.

The purpose of the fake supergravity method is to obtain
workable first order equations whose solutions also satisfy
the Lagrangian equations of motion, but are applicable to
non-BPS solutions of true supergravity theories and to theories
which have only a rough resemblence to supergravity. Even the
limitation to spacetime dimension $D \le 11$ can be overcome in
this framework. The method proceeds by formulating fake Killing
spinor equations whose integrability conditions are the needed first
order equations. One can then attempt to solve these
equations to find new spacetimes or, in combination with the
Witten-Nester approach to gravitational stability,
use it to establish linear stability of previously known
non-BPS solutions.

This approach was first devised for flat-sliced
domain walls in \cite{st} and \cite{modeling} for a bosonic action of
the form 
\be \lab{sdw}
  S_\rom{B} = 
  \int d^{d+1}x \sqrt{-g} 
  \bigg[\half R - \half g^{\m\n}\pa_\m\phi\pa_\n\phi -
V(\phi)\bigg].
\ee
The metric and scalar field of these domain walls take
the form
\bea 
  \lab{flat}
  ds_{d+1}^2 &=& e^{2A(r)}\eta_{\m\n}dx^\m dx^\n + dr^2\, ,
  \\ \nonumber
  \phi &=& \phi(r)\, .
\eea
The warp factor $e^{2A}$ multiplies the metric of $d$-dimensional 
Minkowski spacetime.
The basic quantity in the fake supergravity method is the real
superpotential $W(\phi)$ which is related to the scalar potential by 
\be \lab{wv}
V(\phi) = 2(d-1)^2\Big[W'(\phi)^2 - \frac{d}{d-1}W(\phi)^2\Big].
\ee
The first order flow equations obtained in \cite{st,modeling}, namely
\bea \lab{fofe}
\phi'(r) &=& -2(d-1)W'(\phi) \, ,\nonumber \\
A'(r) &=& 2 W(\phi(r))\, ,
\eea
were later shown in \cite{dbvv}  to be Hamilton-Jacobi equations for
the domain wall dynamics obtained from the field equations of
\reef{sdw}, in which $W(\phi)$ is Hamilton's principal function. The
fake supergravity (or Hamilton-Jacobi) method has had several
applications, especially to brane-world models
\cite{modeling,mannheim,Csaki:2000fc} with stabilized inter-brane
spacing. 

The fake supergravity method works, and is far less restrictive
than true supergravity, because it requires the general structure of
supergravity only to lowest order in fermion fields. Specifically,
as we show in section \ref{s:fakesugra}, one can find a fermion action
$S_\rom{F}$, strictly 
bilinear in the gravitino and dilatino fields $\psi_\m$ and
$\lambda$, such that the sum $S_\rom{B}+S_\rom{F}$ is invariant under local
supersymmetry, but only to linear order in $\psi_\m$ and $\lambda$. To
this order, one requires detailed $\g$-matrix algebra, but
dimension-specific properties such as Fierz rearrangement are not
used. The fake Killing spinor equations are the conditions
 $\d\psi_\m=0$ and
$\d\lambda=0$ obtained from the fermion variations used to
demonstrate linear local supersymmetry.

The next step in the development was the extension of the method to
AdS$_d$ sliced domain walls \cite{fake}. The new metric ansatz replaces
the Minkowski metric $\eta_{\m\n}$ in \reef{flat} with an AdS$_d$
metric  $g_{\m\n}$.
The fake supergravity framework for flat-sliced walls
must be modified because the Lagrangian equations of motion change.
The needed modification incorporates a feature of true $D=5,~\cn=2$
supergravity, namely that the scalar superpotential $W$ is
replaced by an $SU(2)$ matrix\footnote{It is interesting to ask
  whether there is a modification of the Hamilton-Jacobi method in
  which the dynamics is encoded in 
such a matrix.} $\bf{W}$ subject to a further constraint 
reviewed in section \ref{s:AdSsliced}. This modification was
applied \cite{fake} to the stability problem of the Janus solution
\cite{janus} of $D=10$ Type IIB supergravity. The structure of fake
supergravity was further studied in \cite{fakeness}.

In this paper we extend the fake supergravity method to
$\mathbb{R}\times S^3$-sliced domain walls in a 5-dimensional
bulk. Our motivation
is to explore the AdS/CFT correspondence for the situation
of the boundary gauge theory on $\mathbb{R}\times S^3$. Many
recent applications of AdS/CFT involve the $D=4~\cn=4$ SYM
theory on this manifold. 

The bosonic action which governs our system is
\be 
  \lab{rs3}
  S_\rom{B}=\int d^5x\, \sqrt{-g} \bigg[\half R 
 -\half g^{\m\n}\pa_\m\phi\pa_\n\phi 
 - \frac14 Q(\phi) F_{\mu\nu}F^{\m\n} - V(\phi) \bigg] \, .
\ee
It includes an abelian gauge field $A_\m$ with non-minimal coupling to 
the scalar $\phi$.  
In section \ref{s:fakegauge} we construct actions $S_\rom{F}$ and
$S_\rom{gauge}$ quadratic in the gravitino and dilatino fields
$\psi_\m$ and $\lambda$  
such that the total action $S=S_\rom{B}+S_\rom{F}+S_\rom{gauge}$ is 
invariant to linear order in the fermions under local 
supersymmetry transformations. These are motivated by the structure of
real 5D supergravity. A main consequence of linear local supersymmetry
is that the function $Q$ and the superpotential $W$ are required to
take the specific form 
\bea
  \lab{QW}
  Q(\phi) = e^{2k \phi} \, ,
  \hspace{1cm}
  W(\phi)  = w_1\, e^{-k \phi} + w_2\, e^{\frac{2}{3k}\phi} \, ,
\eea
where $w_i$ are constants of integrations. 
Thus the only freedom is the choice of the
constant $k$. This is quite different from the previously studied fake
  supersymmetric actions which admitted arbitrary superpotentials in
  the absence of the gauge field.  
The scalar potential resulting from $W$ in \reef{QW} via \reef{wv} has
  a local maximum. Scalar fluctuations around this local maximum have
  mass$^2$ saturating the Breitenlohner-Freedman (BF) bound \cite{bf}
  for all $k$. The bulk 
  scalar $\phi$ approaches the local maximum at the AdS boundary of
  all our solutions, and is therefore dual to a $\Delta =2$ boundary
  gauge theory operator.

{}For $\mathbb{R}\times S^3$-slicings the gauge field is
necessary for non-trivial solutions of the first order equations.
We impose a static solution ansatz which preserves spherical symmetry
and includes only an electric component of the gauge field. This
leaves four functions to be solved for as functions of a radial
coordinate $r$: the scalar $\phi(r)$ field, the gauge potential $A_t =
a(r)$, and two functions $A(r)$ and $B(r)$ which are warp
factors in the metric. 
The fake supersymmetry transformations of the gravitino and
dilatino yield fake Killing spinor conditions $\d\psi_\m=0$ and
$\d\lambda=0$. Their integrability conditions give rise to first order
``flow'' equations for the four functions $A$, $B$, $\phi$, $a$ of our
ansatz (section \ref{s:inte}). 

The first order equations can be solved analytically; we show how in
section \ref{s:sols}.  
The solutions of the flow equations are fake BPS in the sense that
they admit fake Killing spinors. 
The electrically charged solutions are all
nakedly singular, but a non-extremality parameter $\mu$ can be
introduced to hide the singularity behind
an event horizon. The general non-extremal solutions can then be
written  
\bea
  \lab{introsoln}
  ds_5^2 &=& -H^{-2p} \, f\, dt^2 +
  H^{p}\, (f^{-1}\, dy^2 + y^2 d\Omega_{3}^2) \nonumber\\[2mm]
A_t &=& -\frac{\tilde{q}}{q}\sqrt{\frac{3}{2+3k^2}}\,\big(H^{-1}-1\big)\\
  e^{\frac{2}{3k}\phi} &=& H^p \, , \nonumber
\eea
with
\be
  H(y) = 1+ \frac{q}{y^2}\, , ~~~~~~~~~
   f(y) = 1 +\frac{y^2}{L^2}H^{3p} - \frac{\mu}{y^2}
\ee
and
\bea
  p=\frac{2}{2+3k^2}  \, ,
  ~~~~~
  \tilde{q}^2 = q(q+\mu) \, .
\eea
Asymptotically ($y \to \infty$) the solutions approach global
   AdS$_5$. The parameter $\tilde{q}$ is proportional to the electric charge. 

Various special cases of the solutions \reef{introsoln} had 
previously appeared in the literature
\cite{london,Behrndt:1998ns,cvet,myers,cejm}. Most notably, the fake
BPS solutions (which have $\mu=0$) are truly supersymmetric for the
special values of  
$k = 0, \, 1/\sqrt{3}, \, 2/\sqrt{3}$ as they then arise as solutions
of consistently truncated  
$D=5,~\cn=2$ gauged $U(1)^3$ supergravity theory (see for instance
\cite{duff}).  
The Schwarzschild type mass term $\mu$ was added in
\cite{cvet} providing a horizon, and thus giving regular non-BPS
charged spherically symmetric black holes. (In AdS$_5$, regular BPS
black holes carry non-vanishing angular momentum \cite{reall}.)   
The type IIB lift of the solutions was given in
\cite{cejm,Cveticetal}, and later 
interpreted in \cite{myers} as ``superstars'' describing  continuous
distributions of giant gravitons. 

After we found the solutions \reef{introsoln} for general $k$, we
learned that they had been constructed earlier by Gao and Zhang
\cite{gao} who worked with the
second order Lagrangian field equations. Fake supergravity gives
some insight into the structure of the scalar potential. Here we also 
analyze bulk and AdS/CFT properties of the solutions.

In the context of the AdS/CFT correspondence, previously studied
domain wall solutions and their generalizations had interpretations as
gravitational duals of renormalization group flows, and a holographic
$c$-theorem was derived \cite{cthm,cthm2}. Motivated by this, we
construct here a $c$-function which is monotonically decreasing as the
scalar flows from the asymptotic AdS boundary to the interior. This
result relies only on the structure of the field equations. 

We compute in section \ref{s:mass} the holographic stress tensor from
which we derive the mass of the system.  
All fields of the solutions \reef{introsoln} approach the boundary at
their ``vev rate''. For $q > 0$, the extremal solution should be
the gravity dual of an excited state of the boundary gauge theory with 
non-vanishing charge and vev for a scalar operator with $\D=2$. When
the solutions have horizons, we have the dual of an ensemble of such
states at fixed temperature. Since the gauge theory is on the compact
domain $S^3$, the charge is that of a global symmetry. For the
$k$-values in which the solutions coincide with the superstars of
\cite{myers}, this is an $SO(2)$ subgroup of the $SO(6)$ $R$-symmetry
of $d=4~\cn=4$ SYM theory.

The mass obtained from the holographic stress tensor is suggestive of
a BPS bound saturated for the $\mu=0$ solutions. We compute the
Witten-Nester energy for all solutions, but find that despite the
existence of fake BPS Killing spinors there is an obstruction to
deriving a fully general BPS bound for all $k$. Restricting to the
class of solutions for which $F \wedge F$ vanishes, however, allow us
to confirm the bound suggested by the holographic mass calculation. 

Section \ref{s:disc} contains a brief discussion. 
The main paper is concerned with fake supergravity in $D=5$, but
appendix \ref{app:fake} provides details of the derivation of linear
supersymmetry for general dimensions $D=d+1 \ge 4$. For all $k$, the
bulk scalar is dual to a putative boundary theory operator of
dimension $\Delta=d-2$, which is the dimension of a scalar mass
operator. Appendix \ref{app:horizon} analyzes conditions for the
existence of horizons, and appendix \ref{s:fakeKS} constructs fake
Killing spinors for the fake BPS solutions.


\setcounter{equation}{0}
\section{Basics of fake supergravity}
\label{s:fakesugra}

We introduce the basic structure of fake supergravity and present as
examples the construction of flat- and AdS-sliced domain wall
solutions. 

\subsection{Structure of real and fake supergravity}
Fake supergravity shares the structure of the Lagrangian and
transformation rules of supergravity, but requires local
supersymmetry only to linear order in fermion fields. Linear local
supersymmetry allows more freedom in the bosonic sector, even the
freedom of arbitrary spacetime dimension.

To see why this works, consider a generic true supergravity theory
with a collection of boson and fermion fields $B(x)$ and $F(x)$ and 
transformation rules which involve arbitrary spinor parameters
$\e(x)$. The action $S[B,F]$ is locally supersymmetric, which
means that the supersymmetry variation
\be \lab{locsu}
\d S = \int d^Dx \bigg(\frac{\d \Lag}{\d B} \d B 
  + \frac{\d \Lag}{\d F} \d F\bigg) \equiv 0
\ee
vanishes \emph{identically}, for \emph{all} configurations of $B(x),
F(x), \e(x)$. In particular, the terms of each order in
$F$ vanish \emph{independently}. To lowest order, with fermions more
specifically 
described as gravitinos $\psi_\m(x)$ and Dirac spinors $\l (x)$, the
fermion transformations have the generic structure
\bea
(\d B)_0 &=& \bar{\e}\, \G F 
~=~\bar{\e}\, (\G^\m\,  \psi_\m + \G' \, B \, \lambda) \\[2mm]
(\d F)_0 &=& \Bigg\{
\begin{array}{rcl}
(\d\psi_{\m})_0 &=& (D_\m + \G''_{~\mu}\,  B)\e\\[1mm]
(\d\l)_0 &=& (\G^\m\,  \pa_\m B + \G'''\, B )\e. 
\end{array}
\eea
The $\G, \G'$, etc.~are matrices of the Clifford algebra with the
appropriate tensor structure. 

The lowest order term in $\d S$ is \emph{linear} in the fermions; it
takes the form
\be 
  \lab{linsu}
  (\d S)_\rom{lin} =\int d^Dx\,  
  \bigg[\frac{\d \Lag}{\d B} \, (\bar{\e}\, \G F) 
  + \frac{\d \Lag}{\d F}\, (\d F)_0 \bigg] \equiv 0.
\ee
The variation $\d \Lag/\d B$ is purely bosonic to this order, and 
$\d \Lag/\d F$ is 
linear in fermions. Note that $(\d S)_\rom{lin}$ still vanishes for all
configurations of $B(x), F(x), \e(x)$. If $\e$ is a Killing spinor,
then, by definition $(\d F)_0 =0$, and \reef{linsu} then reads
\be \lab{linsu2}
  (\d S)_\rom{lin} =
  \int d^Dx\; \frac{\d \Lag}{\d B}\, (\bar{\e}\, \G F) \,=\,0.
\ee
It vanishes for all configurations of $B(x)$ which support Killing
spinors and all fermion configurations $F(x)$. Thus the sum over all
independent boson fields $B_I(x)$ vanishes locally, viz.
\be
\label{Beqns}
\sum_I \frac{\d \Lag}{\d B_I}\,(\bar{\e}\, \G F)_I =0.
\ee
If the fermion variations $(\bar{\e}\, \G F)_I$ are independent, then
  each boson equation of motion $\d \Lag/\d B_I =0$ is satisfied separately.

In many cases the fermion variations are independent, in other cases
one must supplement the equations \reef{Beqns} with gauge field
equations of motion \cite{gauntlett}. It is in this way that a
bosonic field configuration $B_I(x)$ which supports Killing spinors 
can give a solution of the bosonic equations of motion of the theory. 
The first order equations which determine these BPS configurations of
$B_I(x)$ are 
the integrability conditions for the Killing spinor equations
$(\d\psi_{\m})_0 =0$ and $(\d \l)_0 =0$.

To see more specifically how fake supergravity imitates and extends this
result, we construct the linear supergravity for the bosonic action
\be \lab{sdw2}
  S_\rom{B} = 
  \int d^{D}x \sqrt{-g} 
  \bigg[\half R - \half g^{\m\n}\pa_\m\phi\pa_\n\phi -
V(\phi)\bigg].
\ee
We do this in some detail because the construction seems
to be new and is an independent sector of the more general situation with
gauge field.  We consider the action $S= S_\rom{B} + S_\rom{F}$ where
$S_\rom{F}$ is 
strictly bilinear in the supersymmetry partners $\psi_\m$ and $\l$ of the
bosons. $S_\rom{F}$ contains all fermion bilinears suggested by true
supergravity, each with an unknown function of $\phi$ as
coefficient, viz.
\bea \lab{sf1}
  S_\rom{F} &=&\int d^Dx\, \sqrt{-g} \,
  \Big[ 4\,\bar{\psi}_\m \G^{\m\n\r}D_\n\psi_\r 
  + \bar{\l}\G^\m D_\m\l 
  -A(\phi)\,\bar{\l} \l
  -B(\phi)\,\bar{\psi_\m}\G^{\m\n}\psi_\n \nonumber \\[1mm]
  && \hspace{2.5cm}
  -\pa_\n\phi\, ( \bar{\psi}_\m \G^\n \G^\m \l
  -\bar{\l}\,\G^\m \G^\n  \, \psi_\m )
  -  C(\phi)(\bar{\psi}_\m\G^\m\l -\bar{\l}\,\G^\m\psi_\m) \Big] 
  \, . ~~~~~~
\eea
The accompanying linearized transformation rules are
\bea 
  \lab{trf1}
  \begin{array}{rclcrcl}
  \d \psi_\m &=& \big(D_\m + \G_\m W(\phi)\big) \e \, , &~~~&
  \d\l &=& \big(\,\G^\m \pa_\m \phi - E(\phi) \big)\e \, , \\[3mm]
  \d e^a_\m &=& 
  -2\big(\bar{\e}\g^a\psi_\m -\bar{\psi_\m}\g^a\e\big) \, , &&
  \d\phi &=& -\bar{\e}\l -\bar{\l}\e \, .
  \end{array}
\eea
The derivative $D_\mu$ includes the spin-connection,
\bea
  \lab{nabla}
  D_\m = \nabla_\m \equiv \pa_\m + \frac14 \o_{\m ab}\g^{ab} \, .
\eea
\emph{A note on conventions}: we use upper case $\G$ for
  gamma-matrices with coordinate indices, and lower case $\g$ for
  gamma-matrices with frame indices. We use
  $\bar{\psi}= i \psi^\dagger \g^t$.

With some work,\footnote{More specifically, one finds linear
  conditions relating the unknown functions by requiring that the 
  coefficients of the following terms in $\d(S_\rom{B} + S_\rom{F})$ each
  vanish: $(\bar{\l}\G\cdot D\e)$, 
  $(\bar{\l}\G\cdot\pa\phi\,\e)$, 
  $(\bar{\psi}_\m\G^{\m\n}D_\n\e)$, 
  $(\bar{\psi}_\m\pa^\m\phi\e)$ and 
  $(\bar{\psi}_\m\G^{\m\n}\pa_\n\phi\,\e)$, and quadratic
  conditions from the coefficients of $(\bar{\l}\e)$ and
  $(\bar{\psi}_\m\G^\m\e)$.
  The relation \reef{wv} appears as the coefficient of
  $(\bar{\psi}_\m\G^\m\e)$}
  one can show that the local supersymmetry variation $\d( S_\rom{B}
  +S_\rom{F})$ 
  vanishes, provided that the potential $V(\phi)$ is related to the
  superpotential $W(\phi)$ by \reef{wv}, and the unspecified functions
  of the  ansatz \reef{sf1}-\reef{trf1} are given by
\be  
\lab{res1}
  A = -(d-1)\big(2W''-W\big)\, ,~~~~~~ 
  B = 4(d-1)W\, ,~~~~~~ 
  C = E = 2(d-1)W'\, .~
\ee

The computations needed to prove linear local supersymmetry are
similar to those of the component approach to supergravity. They
require considerable $\g$-matrix algebra, but dimension specific 
manipulations, such as Fierz rearrangement, are not required at
linear order. This is the reason that linear local supersymmetry is
valid for any dimension!
One further difference is that in fake supergravity it is not
necessary to specify the class of spinor, e.g. symplectic Majorana
spinors in true $D=5$ supergravity. In our computations we assume that
all spinors are complex Dirac spinors.

The derivation of the linear fake supergravity  action
and transformation rules depends only on the bosonic fields one
is working with, in this case the metric $g_{\m\n}$ and a
single scalar $\phi$. It does not depend on the symmetries of the
solution which is sought. In the next stage of the program one uses
the fermion transformation rules of \reef{trf1} as fake Killing 
spinor equations and explores their integrability conditions
in spacetimes of specific symmetry, such as  flat- and AdS-sliced 
domain walls. These examples are described briefly below.

The construction of the linear local supersymmetry theory can be
bypassed as was done for flat- or AdS-sliced domain walls in
\cite{st,modeling,fake}. 
In the more complicated case of $\mathbb{R}\times S^3$ slicing, we
found it useful to work in two stages, first to construct the linear 
supergravity model and then study the resulting fake Killing
spinor conditions obtained from the model.


\subsection{Flat sliced domain walls}
The metric and scalar field of these domain walls take
the form
\bea 
  \lab{flat2}
  ds_{d+1}^2 &=& e^{2A(r)}\, \eta_{\m\n}\, dx^\m dx^\n + dr^2\, ,\\ \nonumber
\phi &=& \phi(r).
\eea
The warp factor $e^{2A}$ multiplies the metric of $d$-dimensional 
Minkowski spacetime. When the ansatz \reef{flat2} is inserted in the
fake Killing spinor 
conditions (see \reef{trf1}-\reef{res1}), they reduce to
\bea 
  \lab{kilf1}
  \mathcal{D}_r \eps &=& \d\psi_r \; 
      = \; \Big(\pa_r +\g_r W\Big)\e \; = \; 0 \, , \\ 
  \lab{kilf2}
  \mathcal{D}_i \eps &=& \d\psi_i \; 
      = \; \Big(\pa_i - \half A' \g^i\g^r +\g^i\, W\Big)\e \; = \; 0 \, , \\ 
  \lab{kilf3}
  \hat{\mathcal{D}}  \eps &=& \d\l \; = \; \Big(\g^r \phi' -2(d-1)W'\Big)\e \; = \; 0 \, . 
\eea
The condition  $\hat{\mathcal{D}}  \eps = 0$ implies $\phi'^2 =
4(d-1)^2{W'}^2$. Consistency of \reef{kilf1} and \reef{kilf3} requires   
\bea
  [ \mathcal{D}_i,\hat{\mathcal{D}}]\eps
  = -\g^i \Big( A' \phi' - 2 W \phi'\, \g^r \Big)  \eps = 0 \, , 
\eea
which by \reef{kilf3} implies $A' \phi' = - 4(d-1) W  W'$. 
Choosing a definite sign for $\phi'$ we can now 
summarize the  first order flow conditions
\bea \lab{fofe2}
\phi'(r) &=& -2(d-1)W'(\phi) \, ,\nonumber \\[1mm]
A'(r) &=& 2 \, W\big(\phi(r)\big) \, .
\eea
These equations are easily integrated
and solve the field equations
\bea \lab{leflat}
d(d-1) \, A'^2 &=& \big(\phi'^2 -V(\phi)\big) \, ,\\[1mm]
\phi'' +d\, A'\phi' &=& \frac{\pa V}{\pa\phi},
\eea
which are the independent equations obtained from the Einstein
and scalar field equations within the ansatz \reef{flat2}. 
The relationship \reef{wv} between the potential $V$ and 
the superpotential $W$ is reproduced by \reef{leflat} using
\reef{fofe2}. The Killing spinors take the form $\e = e^{A/2}\eta$
where $\eta$ is a constant spinor which satisfies 
$\g^r\eta = - \eta$. 


\subsection{AdS$_d$ sliced domain walls}
\label{s:AdSsliced}

The equations \reef{leflat} are modified for the solution ansatz
\be 
  \lab{mads}
  ds_{d+1}^2 = e^{2A(r)}g_{ij}(x)dx^\m dx^\n + dr^2,
\ee
where $g_{ij}$ is an AdS$_d$ metric with scale $L_d$, 
by the addition of the term $-e^{-2A}/L_d^2$ on the
right side of the $A'^2$ equation \reef{leflat}. The fake Killing
conditions \reef{kilf2} are also modified, namely $\pa_i$ is replaced
by the 
AdS$_d$ covariant derivative $\nabla_i^{\rom{AdS}_d}$ in $\d \psi_i$.  
Following the analysis of \cite{fake}, one finds (from
(4.11) of \cite{fake}) that the integrability conditions are
\emph{inconsistent}. 

Although it is not obvious, the inconsistency can be cured by
doubling the spinors and postulating a matrix superpotential 
$\mathbf{W} = \s^a \, W_a(\phi)$ 
(or equivalently a 3-vector $W_a$). 
The $\s^a$ are the Pauli matrices. 
The matrix $\mathbf{W}$ must satisfy the commutator condition
\be \lab{com}
\big[ \mathbf{W}', (d-1)\mathbf{W}'' +\mathbf{W}\big]\,=\,0.
\ee
If this condition is satisfied then the fake Killing conditions are
consistent, and any solution of the flow equations
\bea \lab{fads}
\phi' &=& 2(d-1)\sqrt{W_a W_a} \, , \\ 
\lab{AdSeA}
e^{-2A} &=& 4 L_d^2\frac{(W_aW_a)(W'_b W'_b) -(W_a W'_a)^2}{(W'_b
W'_b)} \, ,
\eea
produces a solution of the Lagrangian equations of motion. Note that
the second equation is algebraic. When $W_a$ and $W'_a$ are parallel
vectors, the inconsistency referred to above is visible in
\reef{AdSeA}. See \cite{fake} for further details of the analysis.

It was not necessary to construct a linear fake supergravity model in
\cite{fake}, but it is quite easy to do so as outlined above. After
doubling all spinors and including ${\bf W}$ one finds that little
new is required. The commutator condition \reef{com} emerges as a
condition for linear supersymmetry. The relation \reef{wv} between $V$
and ${\bf{W}}$ changes only by the replacements $W^2 \to W_aW_a$ and 
$W'^2 \to W'_b W'_b$.

\subsection{$\mathbb{R} \times S^3$ solutions}
\label{s:S3soln}

With $\mathbb{R} \times S^3$-slicing, the metric and scalar fields
take the form
\bea 
  \nonumber
  ds_5^2 &=&- e^{2A(r)}dt^2 + dr^2 + e^{2B(r)}d\Omega_3^2\, ,
  \hspace{1.5cm}
  \phi~=~\phi(r) \, .
\eea
Because of the positive curvature of $S^3$, the Killing spinor
  equations \reef{trf1}-\reef{res1} allow only pure AdS$_5$ as a
  solution, even with a matrix superpotential. To obtain more general
  solutions we add a gauge field to the system, as we discuss in the
  next section. 


\setcounter{equation}{0}
\section{Fake supergravity with a gauge field}
\label{s:fakegauge}

We first describe how to modify the fake supergravity action to
include a gauge field coupled to the scalar, then study the
equations of motion and a $c$-theorem.

\subsection{Fake supergravity action}

In this section we outline the construction of the fake supergravity
model associated with the bosonic action 
\be \lab{action}
  S_\rom{B} =\int d^Dx\, \sqrt{-g} \bigg[\, \half R 
  -\half g^{\m\n} \pa_\m\phi \, \pa_\n\phi 
  - \frac14 Q(\phi) F_{\mu\nu}F^{\m\n} - V(\phi) \bigg] \, .
\ee
It is a considerable complication to add
the gauge field to the previous model specified by
\reef{sdw2}, \reef{sf1} and \reef{trf1}. 
We need to construct an additional bilinear fermion action
$S_\rom{gauge}$ and transformation rules so that the variation of the
total action 
\bea
  \lab{total}
  S = S_\rom{B} + S_\rom{F} + S_\rom{gauge}
\eea
vanishes to linear order in $\psi_\m$ and $\l$. $S_\rom{B}$ is
given in \reef{action} and $S_\rom{F}$ in \reef{sf1}, and we take for 
$S_\rom{gauge}$
\bea 
  \lab{sga}
   S_{{\rm gauge}} &=& \int d^Dx \sqrt{-g}\, 
  \Big[-i\,M(\phi)\,\bar{\psi}_\m\big(\G^\m\G^{\r\s}\G^\n 
  -\G^\n\G^{\r\s}\G^\m\big)\psi_\n F_{\r\s}
  + i\,P(\phi)\,\bar{\l}\,\G^{\r\s}F_{\r\s}\l~~\nonumber \\
  &&\hspace{2.5cm}
  + i\,N(\phi)\,\big( \bar{\psi_\m}\G^{\r\s} \G^\m\l
  -\bar{\l}\,\G^\m\G^{\r\s}\psi_\m \big) F_{\r\s}\Big].
\eea
Each term in $S_{{\rm gauge}}$ consists of a fermion bilinear with
   the same $\g$-matrix structure as in true supergravity
multiplied by a function of $\phi$ to be determined. 
The gravitino $\psi_\m$ and the dilatino $\l$ are charged, hence
the derivatives $D_\m$ that appear in the two fermion kinetic terms of
\reef{sf1} now include a coupling to the gauge field,
\be 
  \lab{cov}
  D_\m = \nabla_\m + ic A_\m \, , 
\ee
where $\nabla_\m$ as defined in \reef{nabla} contains the spin
connection. Gauge invariance
requires that the gravitino and dilatino carry the same charge so that
the mixed $\bar{\lambda} (\cdots) \psi_\m$ terms in \reef{sf1} are
gauge invariant. The scalar $\phi$ is neutral. 

We also postulate the following transformation rules
\bea 
  \lab{trf2}
  \d\psi_\m &=&
   \big[D_\m +\G_\m W(\phi) 
    +iX(\phi)\big(\G_\m\,^{\n\r}-2(D-3)\, 
      \d_\m^\n\, \G^\r\big) F_{\n\r} 
    + ic A_\m\big]\e\, ,\\[1mm] \nonumber
  \d\l &=& \big[\G^\m\, \pa_\m\phi - 2(D-2)W'(\phi) 
    +i\,Y(\phi)\, \G^{\r\s}F_{\r\s} \big]\e\, ,\\[1mm] \nonumber
  \d e^a_\m &=& 
    -2\big(\bar{\e}\g^a\psi_\m -\bar{\psi_\m}\g^a\e\big)\, ,\\[1mm] 
  \nonumber
  \d\phi &=& -\bar{\e}\l -\bar{\l}\e \, ,\\[1mm] \nonumber
  \d A_\m &=& -i\a(\phi) 
    \big(\bar{\e}\,\psi_\m - \bar{\psi}_\m \e\big)
    -i\b(\phi)\big(\bar{\e}\, \G_\m\l+\bar{\l}\, \G_\m\e\big)\, .
\eea

The requirement of linear local supersymmetry of the total action
\reef{total} fixes all unknown scalar functions and the
$U(1)$ coupling $c$. Terms independent of the gauge field are a
closed sector of the calculation, so the results \reef{res1}
for $A,\,B,\,C,\,E$ remain valid.
To extend linear local supersymmetry to the gauge sector, we need
to examine about 16 distinct spinor bilinears. The coefficient of
each is a combination of the unspecified scalar functions of the
ansatz in \reef{sga} and \reef{trf2} and derivatives of those
functions. Each such combination must vanish. The information in
these conditions fixes the scalar functions uniquely up to
integration constants which we then specify by imposing physical
normalization conditions. The analysis of the 16 conditions is
tedious, so we simply quote results for $D=5$ here. 
Further details for general $D$ are given in appendix \ref{app:fake}.

The results for $Q$, $X$, $Y$, $W$, and $c$ which are actually needed 
to study the fake Killing spinor conditions are:
\bea 
  \lab{qxyw}
  Q(\phi) &=& e^{2k\phi}\, , \hspace{8mm}
  X(\phi) ~=~ \frac{1}{4\sqrt{3(2+3k^2)}}\, e^{k\phi}\, , \hspace{8mm}
  Y(\phi) ~=~ 6k\,  X(\phi)\, , \\[2mm]\nonumber
  W(\phi) &=& \frac{1}{2L(2+3k^2)}\Big(2\, e^{-k\phi} + 3k^2
              \, e^{\frac{2\phi}{3k}}\Big)\, , \hspace{8mm}
  c ~=~ -\frac{1}{L} \sqrt{\frac{3}{2+3k^2}} \, ,
\eea
while scalar functions in the actions \reef{sf1} and
\reef{sga} and the boson transformation rules of \reef{trf2} are
\bea
  \begin{array}{rclcrclcrcl}
  M(\phi) &=& -6X(\phi)\, , &&
  N(\phi) &=& Y(\phi)\, , &&
  P(\phi) &=& 3(1-2k^2)X(\phi) \, , \\[3mm] 
  \a(\phi)&=& -24\,\displaystyle{\frac{X(\phi)}{Q(\phi)}}\, ,&&
  \b(\phi)&=&12k\,\displaystyle{\frac{X(\phi)}{Q(\phi)}} \, .
  \end{array}
\eea

The potential $V(\phi)$, obtained by inserting the superpotential
from \reef{qxyw} into \reef{wv}, has a unique local maximum at $\phi
=0$. 
This is the asymptotic value of the scalar in all the solutions we
obtain. 
It is straightforward to expand the potential near
the maximum and compare with the potential of a general
massive scalar in AdS$_5$ of scale $L$:
\bea
V(\phi) &=& -\frac{6}{L^2} -\frac{2}{L^2}\phi^2 
  ~=~  -\frac{6}{L^2} +\half m^2\phi^2.
\eea
We see that the parameter $k$ has cancelled and the bulk scalar field 
has mass, $m^2 = -4/L^2$, thus saturating the BF bound \cite{bf}
for all values of $k$. It is curious that fake supergravity, with
one scalar and one gauge field, selects this value.\footnote{As shown
  in appendix \ref{app:fake}, for $D \ne 5$ the mass $m^2$ is strictly
  above the BF bound.} The potential is analyzed further in section
\ref{s:flow}.


\subsection{Equations of motion}

The goal of this paper is to apply fake supergravity methods to
obtain solutions of the equations of motion of the bosonic
theory \reef{action} within the ansatz
\bea 
  \nonumber
  ds_5^2 &=&- e^{2A(r)}dt^2 + e^{2h(r)}dr^2 + e^{2B(r)}d\Omega_3^2 \, ,\\
  \lab{rs3ansatz}
  \phi &=& \phi(r) \, ,\\
  \nonumber
  F_{rt} &=& \pa_r A_t(r) \equiv \Uprime(r)\, .
\eea
The gauge field configuration is purely electric. 
The function
$h(r)$ specifies the choice of radial coordinate, and we keep
this freedom because different coordinates are convenient at different
points in our study. 
It is useful to employ a definite form of the 3-sphere metric,
namely
\bea \lab{3s}
d\O^2_3 &=& d\theta^2 + \sin^2\theta\, d\phi^2
+\cos^2\theta\, d\psi^2\, ,
\eea
with coordinate ranges $\theta \in [0, \frac{\pi}{2}]$ and
$\phi,\psi \in [0, 2\pi]$.

The gravitational equations of motion are
\bea \lab{graveq}
R_{\m\n} &=& T_{\mu\nu} - \frac13 g_{\m\n} T_\r\,^\r\\
  &=& \pa_\m \phi \,\pa_\n\phi + g_{\m\n}\Big(\,\frac23 \,  V
  -\frac{1}{6}\, Q\, F^2\Big)+Q\, F_\m\,^\r F_{\n\r}.
\eea

In the ansatz \reef{rs3ansatz}, these equations become
\bea 
\lab{rr}
R_{rr} &=& -A'' -3B'' +(A'+3B')(h'-A'-B') + 4 A' B' \\ \nonumber
    &=& \phi'^2 +\frac23 e^{2h}\, V 
       - \frac{2}{3} e^{-2A}\, a'^2 \, Q \, , \\[3mm]
\lab{tt}
R_{tt} &=&e^{2A-2h}\big(A''+A'(A'+3B'-h')\big)\\ \nonumber
    &=& - \frac23 e^{2A}\, V + \frac{2}{3} e^{-2h} \,a'^2 \, Q
  \, , \\[3mm]
\lab{thth}
R_{\theta\theta}&=&2- e^{2B-2h}\big( B'' + B'(A' +3B' - h')\big) \\ \nonumber
&=&e^{2B}\Big(\,\frac23 \, V +\frac{1}{3} e^{-2A-2h}\, a'^2 \, Q \Big).
\eea
Note that $R_{\phi\phi}=\sin^2\theta R_{\theta\theta},~
R_{\psi\psi}=\cos^2\theta R_{\theta\theta},$ and that off-diagonal
components of the Ricci and stress tensors vanish.

Later we will use the following combinations of the equations
above:
\bea
\label{Beom}
 e^{-A-3B+h}\Big(e^{A+3B-h}\, B'\Big)' &=&
 -\frac{2}{3} e^{2h}\, V +
 2 e^{2h-2B}
 -\frac{1}{3} e^{-2A} \, \Uprime^2 \, Q \, ,\\
 \label{Aeom}
  e^{-A-3B+h}\Big(e^{A+3B-h}\, A'\Big)'
  &=&
 -\frac{2}{3} e^{2h}\, V
 +\frac{2}{3} e^{-2A} \, \Uprime^2 \, Q \, , \\
  \frac{1}{2}  {\phi'}^2 -3B'^2 - 3 A' B'&=&
  -3 e^{2h-2B}
  +e^{2h}\, V
 +\frac{1}{2} e^{-2A} \, \Uprime^2 \, Q \, .~~~
\label{hEOM}
\eea
The gauge field and scalar equations of motion are
 \bea
 \Big(e^{-A+3B-h}\, Q\, \Uprime\Big)' = 0 \, .
\label{gaugeEOM}
\eea
\bea
 e^{-A-3B+h}\Big(e^{A+3B-h}\, \phi'\Big)' = e^{2h}\, \frac{\pa V}{\pa \phi}
 -\frac{1}{2} e^{-2A} \, \Uprime^2 \,  \frac{\pa Q}{\pa \phi} \, .
\label{scalarEOM}
\eea
The equations of motion can also be obtained from the one-dimensional effective action
\bea
  S&=& -\int dr\, e^{A+3B-h}
  \bigg[
    \frac{1}{2} \phi'^2 - 3B'^2 - 3 A' B'
     -\frac{1}{2} e^{-2A} \, \Uprime^2\,  Q(\phi)
     - 3 e^{2h-2B} +e^{2h}\, V(\phi)
  \bigg] \, . \nonumber\\
\eea

Note that the field equations are not all independent. For
example, ~\reef{Aeom} can be derived by differentiating
\reef{hEOM} and using the other equations of motion.


\subsection{A $c$-theorem}

The combination $R_{rr} + e^{-2(A-h)}R_{tt}$ of the Ricci tensor
components gives
\be \lab{simp}
-3(B'' +B'^2 -A'B' -B'h') = \phi'^2.
\ee
If we choose $h=B-A$ and call the corresponding radial coordinate
$\tilde{r}$, then, with $'$ denoting $d/d\tilde{r}$, we find that
the quantity
\be \lab{cfun}
C(\tilde{r}) \equiv \frac{\cc_0}{B'(\tilde{r})^3},
\ee
is monotonic increasing with $\tilde{r}$ for any positive
constant $\cc_0$. 

We wish to adapt the argument of \cite{cthm,cthm2} and interpret
$C(\tilde{r})$ as a $c$-function.
For this purpose we write the AdS$_5$ metric
using two different radial coordinates, the first corresponds to $h=0$ and
the second is $\tilde{r}$:
\bea \lab{adsr}
ds^2_{\rom{AdS}_5} &=& -L^2 \cosh^2\Big(\frac{r}{L}\Big) \, dt^2 + dr^2 + L^2
\sinh^2\Big(\frac{r}{L}\Big)\, d\O^2_3\\
\lab{adsrt}
&=& -L^2(1+ e^{2\tilde{r}/L}) \, dt^2
+ \frac{d\tilde{r}^2}{1+e^{-2\tilde{r}/L}} 
  +L^2\, e^{2\tilde{r}/L}\, d\O^2_3.
\eea
The two coordinates are related by $\tilde{r} = L \ln(\sinh(r/L))$.
We see that $B'(\tilde{r})=1/L$. Now consider a solution of the
equations of motion in which the $\mathbb{R} \times S^3$ sliced metric
approaches the boundary region, $\tilde{r}\to +\infty$, of an
 AdS$_5$ spacetime with scale $L_\rom{UV}$ and the deep interior region,
$\tilde{r}\to-\infty$, of an AdS$_5$  spacetime with scale $L_\rom{IR}$. The
$c$-function \reef{cfun} then interpolates monotonically between
these limits. With suitable normalization, i.e. choice of $\cc_0$,
it coincides at the endpoints with the central charge
 \cite{hr1} of putative dual 4-dimensional conformal field
theories on $\mathbb{R} \times S^3$. Since $L_\rom{IR} < L_\rom{UV}$,
the central charge decreases in the renormalization group flow toward
long distance.

It would be strange if the construction of a $c$-function required a
particular radial coordinate, and indeed it does not. For any choice
of $h(r)$, it is straightforward to see, using \reef{simp}, that
\be
 C(r) \equiv \cc_0 \bigg(\frac{dB}{dr}\bigg)^{-3} e^{3(h+A-B)}
\ee
is monotonic and in fact coincides with $C(\tilde{r})$. The
interpretation is more straightforward with the $\tilde{r}$
coordinate (and the AdS$_5$ warp factor $e^{\frac{2\tilde{r}}{L}}$
is pure exponential), but the physics is more general. The
monotonicity of $C$ depends only on the equations of motion for the
solution ansatz, not the actual solution. 

The interpretation of the $c$-function will be less clear for our
solutions because they contain a singularity in the interior. 
It turns out that the $c$-function $C(r)$ is
non-vanishing at horizons, when present, while $C(r)$ vanishes at the  
singularity.

\setcounter{equation}{0}
\section{Fake BPS and non-extremal solutions}
\label{s:sols}

We derive integrability conditions from the Killing spinor
conditions of the fake supergravity model of section
\ref{s:fakegauge}.  
We then solve these first order condition
to obtain the most general fake BPS solutions within the ansatz
\reef{rs3ansatz}. The solutions are then generalized to include a
non-extremality parameter. We study relevant properties of the
solutions.


\subsection{Integrability conditions for fake Killing spinors}
\label{s:inte}

The Killing spinor equations obtained from the fermion transformation
rules in \reef{trf2} are
\bea
 \label{Kspin}
 \mathcal{D}_\mu \eps &\equiv&
 \Big[
   D_\mu
    + i X(\phi) \Big(\Gamma_\mu^{~\nu\rho}
      -4\delta_\mu^{~\nu}
      \Gamma^\rho\Big) F_{\nu\rho}
    + \Gamma_\mu W(\phi) + i c\, A_\mu
  \Big] \eps \,=0\, , \\[2mm] \nonumber
\hat{\mathcal{D}} \eps &\equiv&
 \Big[
   \Gamma^\mu \pa_\mu \phi
    -6 W'(\phi)
    + i Y(\phi)  \Gamma^{\mu\nu} F_{\mu\nu}
  \Big] \eps \,=0\, .
\eea
In the obvious diagonal frame for the metric
\reef{rs3ansatz}-\reef{3s}, and with spin connections included, the
operators in \reef{Kspin} become
\bea \lab{kdil}
  \hat{\mathcal{D}} &=&
  e^{-h} \phi'\, \gamma^r
  - 6 W'
  + 2i \Uprime Y e^{-A-h} \gamma^r \gamma^t \, , \\[2mm]
\lab{kt}
\mathcal{D}_t &=& \pa_t
  -\frac{1}{2} A' e^{A-h}  \, \gamma^t \gamma^r
  + 4i  \Uprime X e^{-h} \gamma^r
  - e^A W \gamma^t
  + i c A_t \, , \\[2mm]
\lab{kr}
\mathcal{D}_r &=& \pa_r
   + e^h W \, \gamma^r
   -4i \Uprime X e^{-A} \, \gamma^t \, , \\[2mm]
\lab{kth}
\mathcal{D}_\theta &=& \pa_\theta
   + \frac{1}{2}B' e^{B-h}\, \gamma^\theta \gamma^r
   + e^B W \,\gamma^\theta
   +2i \Uprime X e^{B-A-h} \, \gamma^\theta \gamma^r \gamma^t \, ,\\[2mm]
\lab{kph}
\mathcal{D}_\phi &=& \pa_\phi
   + \frac{1}{2} \, \gamma^\phi \gamma^\theta  \cos\theta
   +\gamma^\phi  \left( \frac{1}{2}B' e^{B-h}\,  \gamma^r
   + e^B W
   +2i \Uprime X e^{B-A-h} \, \gamma^r \gamma^t  \right) \sin\theta\,
   .~~~ \nonumber \\
\eea
Note that $'$ means $d/dr$ for the functions $A$, $B$, $h$, $a$, and
 $\phi$ of the  
solution ansatz of \reef{rs3ansatz}, but means $d/d\phi$ for the
superpotential $W(\phi)$.

Fake Killing spinors are solutions of the equations \reef{Kspin}.
Solutions exist if the commutators of the 6 conditions vanish, i.e.
\bea
  \label{commutators}
  [\mathcal{D}_\mu,\hat{\mathcal{D}} ]\e = 0 \, ,
  ~~~~~
  [\mathcal{D}_\mu,\mathcal{D}_\nu]\e = 0 \, .
\eea
The commutator conditions are a set of first and second order
differential equations for $A(r)$, $B(r)$, $a(r)$, $\phi(r)$. It turns
  out that only the first order conditions, those obtained from
  commutators not 
involving $\mathcal{D}_r$, are sufficient to obtain solutions of the
Lagrangian equations of motion \reef{Beom}-\reef{scalarEOM}. Since
the full analysis is tedious, we simply present some essential
points and the results for the set of four first order scalar
equations which we actually use. In appendix \ref{s:fakeKS}, we will
  present explicit fake Killing spinors which will serve as a check
  that the full set of commutator conditions is satisfied.

The dilatino equation $e^h \g^r\, \hat{\mathcal{D}} \eps=0$ reads
\bea
  (\phi' -6\, e^h W' \,\gamma^r
  + 2i Y \Uprime e^{-A} \,\gamma^t) \eps = 0 \, .
\label{Dproj}
\eea
It is essential to use this constraint on $\e$ to derive the
integrability conditions.  For example, if we multiply \reef{Dproj}
by $(\phi' +6\, e^h W' \,\gamma^r -2i Y \Uprime e^{-A} \,\gamma^t)$,
we obtain the scalar equation
\bea
  \label{dilasq}
  \phi'^2 = 36 W'^2 e^{2h} + 4 Y^2 \Uprime^2 e^{-2A} \, .
\eea
{}From commutators not involving $\mathcal{D}_r$ we obtain the three
additional equations
\bea
 \label{DDth}
  B' W' &=& -\frac{1}{3} W \phi' \\[2mm]
  \label{DDt}
  A' \phi'&=&  - 12 \, e^{2h} W W' - 16 X Y \Uprime^2\, e^{-2A}  \\[2mm]
 \label{3rd}
  A' B' + B'^2 &=&   8 e^{2h}W^2 -  16 X^2 \Uprime^2 e^{-2A}
  +e^{2h-2B}.
 \eea
This system of first order coupled differential equations can be
  solved exactly, as we demonstrate next.
The specific functions $X(\phi)$, $Y(\phi)$, $Q(\phi)$, $W(\phi)$ given
in \reef{qxyw} are used to obtain the solution.


\subsection{Construction}

We start by integrating the gauge equation of motion \reef{gaugeEOM},
and write its square as
\bea
  \label{gaugeint}
  \Uprime^2\, e^{-2A}X^2\, = \s^2 X^{-2} e^{2h-6B}.
\eea
where $\s$ is an integration constant.
This relation may be
inserted in \reef{dilasq} and the four conditions
\reef{dilasq}-\reef{3rd} then reduce to coupled equations for the
metric functions $A$, $B$, $h$. To solve them it is useful to treat $A$,
$B$, $h$ as functions of the scalar $\phi$. Temporarily introducing a
dot to denote derivatives with respect to $\phi$, the equations
become 
\bea
  \label{phieq}
  e^{-2h} \phi'^2 &=& 36 \dot{W}^2
   + 144 k^2 \s^2 X^{-2} e^{-6B} \, , \\
  \label{Bdot}
  \dot{B} &=& -\frac{1}{3} \frac{W}{\dot{W}} \, , \\
  \label{Adot}
  e^{-2h} \phi'^2 \dot{A} &=& -12 W \dot{W}
  - 96 k \s^2 X^{-2} e^{-6B}\, ,  \\
  \label{ABdot}
  e^{-2h} \phi'^2 (\dot{A}\, \dot{B}+\dot{B}^2)
  &=& 8 W^2 -16   \s^2 X^{-2} e^{-6B} + e^{2B}\, .
\eea
Plugging equations \reef{phieq}-\reef{Adot} into the LHS of
eq.~\reef{ABdot}
we find a very simple algebraic equation for $B$,
\bea
  \label{Balg}
  e^{-2B} =  \left|\frac{1}{4\s} X \frac{\dot{W}}{\dot{W}+k W}\right| \, .
\eea
However, eq.~\reef{Bdot} can also be integrated directly using the
superpotential in \reef{qxyw}.
Including a constant of integration, $c_B$, we find
\bea
  \label{Bint}
  e^{-2B} = c_B \left(e^{k\phi} -  e^{-\frac{2}{3k}\phi} \right)
  = \frac{c_B\, (2+3k^2) L}{k}\, e^{k \phi - \frac{2}{3k}\phi} \,
  \dot{W}(\phi) \, .
\eea
The expressions for $e^{-2B(\phi)}$ in \reef{Balg} and \reef{Bint}
are proportional. Requiring equality fixes the relationship between
the two integration constants $\sigma$ and $c_B$,
\bea
  \label{consistency1}
  |\sigma c_B| = \frac{1}{8\sqrt{3 (2+3k^2)^{3}}} \, .
\eea
This can be understood as a condition that the first order equations
\reef{phieq}-\reef{ABdot} are mutually consistent.

Next integrate \reef{Adot} to find
\bea \lab{aint}
  e^{2A} = c_A\,  e^{-\frac{4}{3k} \phi}
  \left[ c_B L^2 + e^{\frac{2}{k}\phi}
    \Big( e^{ \frac{2+3k^2}{3k}\phi}-1 \Big)^{-1}\right] \, .
\eea

It remains to find the scalar profile $\phi(r)$. Although \reef{phieq}
gives a separable equation, it is difficult to integrate, so we
proceed differently.
We note that static 5D black holes in the literature (see \cite{duff} and
references therein) are most simply described via the point singular
harmonic function $H(y) = 1 + q/y^2$, and that $H$ and $\phi$ are
related logarithmically. We introduce $H$ in two stages, first
defining  
\bea
  \label{phi1}
  \phi(H) \equiv \frac{3k}{2+3k^2}  \log{H} \, ,
\eea
where the multiplicative constant was chosen to simplify
\reef{Bint} and \reef{aint}.

We temporararily regard $H$ as the radial coordinate, $r=H$,
which means
that $\phi' = 3k/[(2+3k^2)H]$. Then \reef{phieq}
immediately determines $h$ as a function of $H$:
\bea
  \label{expm2h}
  e^{-2h}&=&  4(2+3k^2)^2\,H^2 \, \left(  \dot{W}^2
   + 4 k^2 \s^2 X^{-2} e^{-6B} \right)  \\ \nonumber
&=& \frac{4}{L^2 H^p} (H-1)^2\Big[H^{3p} + c_B L^2 (H-1)\Big],
\eea
in which we have introduced the constant
\be
  p = \frac{2}{2+3k^2} \, .
\ee
The scale factors of the metric can then be written as
\bea
  e^{2A} &=& c_A  L^2 H^{-2p}
  \left( c_B + \frac{1}{L^2(H-1)} H^{3p} \right) \, ,\\
  e^{2B} &=& \frac{1}{c_B(H-1)} H^p \, ,\\
  e^{2h} &=& \frac{1}{4(H-1)^3} H^p
  \left( c_B + \frac{1}{L^2(H-1)} H^{3p} \right)^{-1} \, .
\eea
The line element now contains the term $e^{2h}dH^2$.

The scale factors $e^{2A}$ and $e^{2B}$ diverge as $H$ approaces
$H=1$, indicating that this is asymptotic infinity (where
we will find an AdS$_5$ boundary). The scalar $\phi \to
0$ in this limit, and $\phi=0$ is the unique root of $W'(\phi)=0$
and is a local maximum of the potential $V(\phi)$. We therefore
introduce the radial coordinate $y$ such that
 $H \to 1$ for $y \to \infty$, i.e.
\bea
  \label{Hofy}
  H(y) =1 + \frac{q}{y^2} \, .
\eea
Here $q$ is a constant of dimension (length)$^2$.

The radial term in the line element now becomes
\bea
  e^{2h}\, dH^2 = e^{2h} (\pa_y H)^2\, dy^2
  = H^p \left( c_B\, q + \frac{y^2}{L^2} H^{3p} \right)^{-1} dy^2
\eea
and, the other scale factors are
\bea
 e^{2A} &=& \frac{c_A L^2}{q} H^{-2p}
  \left( c_B\, q + \frac{y^2}{L^2} H^{3p} \right) \, , \\
 e^{2B} &=& \frac{1}{q\, c_B}\,  y^2 H^p \, .
\eea
The metric of pure AdS$_5$ can be written as
\bea \lab{adsy}
  ds^2 = -\left(1+ \frac{y^2}{L^2}\right) \, dt^2
  + \left(1+\frac{y^2}{L^2}\right)^{-1} \, dy^2
  + y^2 \, d\Omega_3 \,.
\eea
We require that our metric match the leading terms of \reef{adsy}
as $y \to \infty$, and this fixes the remaining integration
constants to be
\bea
  \label{cAcB}
  c_A = \frac{q}{L^2} \, ,
  ~~~~
  c_B = \frac{1}{q} \, .
\eea
Thus we can write the general solution to the first order
equations \reef{dilasq}-\reef{3rd}
\bea \lab{newmet}
  ds^2 = - H^{-2p} f \, dt^2
  + H^p f^{-1} \, dy^2
  + y^2 H^p \, d\Omega_3 \, ,
\eea
where 
\bea \lab{newf}
  H(y) =1 + \frac{q}{y^2}\, ,
  ~~~~~
  f(y) = 1 + \frac{y^2}{L^2}\,  H^{3p} \, .
\eea
(The metric is not conformally flat.)

The scalar is
\bea \lab{newsc}
  e^{\phi} = H^{\frac{3k p}{2}} \, ,
\eea
and the gauge field strength is found from \reef{gaugeint}
\bea \lab{newgf}
  \label{pregaugepot}
  F_{yt} =  \Uprime = \s X^{-2} e^{A-3B+h}
  = 48 (2+3k^2) \, \sigma \, y^{-3} H^{-2} \, .
\eea
Using \reef{consistency1} and \reef{cAcB} we integrate
\reef{pregaugepot} to find
\bea \lab{newat}
  A_t = \UU = \mp \sqrt{\frac{3}{2+3k^2}}\,  \frac{q}{q+y^2}
  = \pm \sqrt{\frac{3}{2+3k^2}} \, (H^{-1}-1)\, ,
\eea
where we have fixed the constant of integration such that $A_t \to 0$
for $y \to \infty$. \footnote{The sign of the gauge potential
and electric field are arbitrary and independent of the sign of
$q$.}

We have constructed the most general solutions of the first order
equations derived from integrability of the
fake Killing spinor equations. We call them fake BPS solutions.
For each value of the parameters $k$ and $L$ from the fake
supergravity action, there is a 1-parameter set of solutions
depending on $q$. The solutions carry electric charge which can
be calculated from the integral
\be \lab{chgint}
q_{\rm elec} = \frac{1}{2\pi^2}\int_{S^3} Q \star F
\ee
over the asymptotic 3-sphere. The result is
\be
 \lab{EXTqelec}
 q_{\rm elec}= \pm 2\sqrt{\frac{3}{2+3k^2}}\,q\,.
\ee
Since there are no charged sources for the gauge field, this
charge is concentrated at the center of the $S^3$, where
the scale factor $e^{2B} = y^2 H^p$ vanishes.


\subsection{Non-extremal solutions}

The solutions constructed above can be generalized beyond
extremality. Inspired by the solutions of
\cite{cvet} we simply modify $f$ and $A_t$ to be
\bea \lab{fmu}
  f(y) = 1 + \frac{y^2}{L^2}\,  H^{3p} - \frac{\mu}{y^2}\,
\eea
and
\bea \lab{amu}
 A_t =
 -\frac{\tilde{q}}{q}\sqrt{\frac{3}{(2+3k^2)}}\,\big(H^{-1}-1\big)\, ,
\eea
where
\bea \lab{tq}
  \tilde{q}^2 = q (q+\mu) \, .
\eea
It is straightforward to verify that the equations of motion
\reef{Beom}-\reef{scalarEOM}  are satisfied for all $k$, but the first order
BPS equations are no longer satisfied. The electric charge is
changed to
\be \lab{elec}
 q_{\rm elec}= \pm 2\sqrt{\frac{3}{2+3k^2}}\,\tilde{q}\,.
\ee


The mass of the solutions with respect to the background AdS$_5$ space
is 
\bea
  M_0 = \frac{\pi}{4G} \left[ \frac{3 \mu}{2}
    + \frac{6}{2+3k^2}\,q \right] \, .
\eea
This is computed in section \ref{s:mass}, where we also discuss a BPS
bound.


\subsection{Comparison with known solutions}
\label{s:comp}

It turns out that none of the extremal solutions found above by
the fake supergravity technique, and none of their non-extremal
extensions, are new. We review relevant past work here beginning with 
solutions found in the AdS/CFT context.

Superpotentials which are the sum of two exponentials, as in
\reef{qxyw}, have occurred before in applications of 5D supergravity,
namely in \cite{fgpw,brand}. There flat sliced domain wall solutions
with no gauge fields were
found. These Coulomb branch solutions lift to type IIB supergravity
and correspond to continuous 
distributions of D3-branes on subspheres of the $S^5$ of dimension
$n=1,\dots,5$. In fact from eq.~(15) of
\cite{fgpw} (after the change 
$\m = \pm \phi/\sqrt{2}$ to agree with our conventions), one can see that
the five superpotentials considered there agree with our $W(\phi)$
for the specific values of $k$
\bea
  \lab{speck}
  \begin{array}{rcrrrrrrrrr}
  k&=&\sqrt{\frac{10}{3}},
   & ~\frac{2}{\sqrt{3}},
   & ~\sqrt{\frac{2}{3}}, 
   & ~\frac{1}{\sqrt{3}},
   & ~\sqrt{\frac{2}{15}},\\[3mm]
  n &=& 1,& 2,&3,&4,&5\, . 
  \end{array}
\eea
There is an even closer relation to our work. Namely, the scale factor 
$A$ of the flat-sliced ansatz \reef{mads}, expressed as a
function of the scalar field as $A(\phi)$ obeys the same equation
\reef{Bdot} as our $B(\phi)$ and has the same solution.

For $k=2/\sqrt{3}$ and $1/\sqrt{3}$ the theory
\reef{action} can be recognized as special cases of the supersymmetric  
$U(1)^3$ theory which again is a consistent truncation of Type IIB
supergravity on $S^5$. The $U(1)^3$ theory consistently truncated to a
single scalar field includes two gauge fields \cite{duff}. The two
solutions with $k=2/\sqrt{3}$ and $1/\sqrt{3}$ correspond to setting
either of those gauge fields to zero.\footnote{This is not a
  consistent truncation for $k=1/\sqrt{3}$, because the gauge
  field set to zero is then sourced by a potentially non-vanishing $F
  \wedge F$ term. Note that in our
  analysis of linear supergravity it was in fact only the $k=0$ and
  $k=2/\sqrt{3}$ cases which were fully linearly supersymmetric (see
  appendix \ref{app:fake} for details).} 
Solutions of the further
truncated theory, for which the scalars completely decouple, leaving 5D
minimal gauged supergravity, can be obtained as the 
$k \to 0$ limit (appropriately defined) of our solutions.
Since they can be embedded as solutions of the supersymmetric $U(1)^3$
theory, the fake BPS solutions are in the three cases, 
$k=0,\, 1/\sqrt{3},\, 2/\sqrt{3}$, truly 
supersymmetric \cite{london,Behrndt:1998ns,myers}. 
Their non-extremal
generalizations coincide with those of \cite{cvet}.   
The 10D lifts \cite{cejm,Cveticetal} of these solutions, known as
``superstars'', describe distributions of giant gravitons rotating on
the $S^5$ \cite{myers}.

One might hope to lift the other three values of $k$ from table
\reef{speck}. This requires going beyond the $U(1)^3$ truncation, for
example to the gauged $SO(6)$ truncation
\cite{Cvetic:2000nc}. However, it appears that the gauge kinetic
function of \cite{Cvetic:2000nc} is not compatible with our $Q(\phi)$
for the relevant values of $k$ \cite{michael}. 
It seems unlikely that the solutions can be lifted to 10D for general
$k$. 

The extremal and non-extremal solutions for general $k$ are not new, but were
found in \cite{gao}.\footnote{The solutions of \cite{gao} are
presented using a different coordinate system; it is easy to relate
their choice of radial coordinate to our $y$.} Ref.~\cite{gao}
constructed similar solutions for any $D \ge 4$. For general $D \ge 4$
the scalar potential of \cite{gao} can indeed be constructed
from our superpotential \reef{supot}. 


\subsection{Scalar ``flow'' and horizons}
\label{s:flow}

The scalar potential \reef{wv} constructed from the superpotential
\reef{qxyw} is
\bea
  \nonumber
  V(\phi) &=& 
  -\frac{6}{(2+3k^2)^2\, L^2} 
  \bigg[ 
    18 k^2 \,
    e^{\big( \frac{2}{3k}-k\big)\, \phi} 
    +3k^2(3k^2-1) \,
    e^{ \frac{4}{3k}\, \phi}
    -(3k^2-4)\, 
    e^{-2 k \,  \phi}
  \bigg]\, . \\
\eea
The behavior of $V$ depends on the 
value\footnote{We restrict to $k>0$, since $k \to -k$ is equivalent to 
  taking $\phi \to - \phi$.} of $0<k<\infty$, but in all cases there
is a local maximum at $\phi=0$, which occurs at the AdS boundary of the
solutions. 

\begin{figure}[t]
\centerline{
 \includegraphics[width=4.3cm]{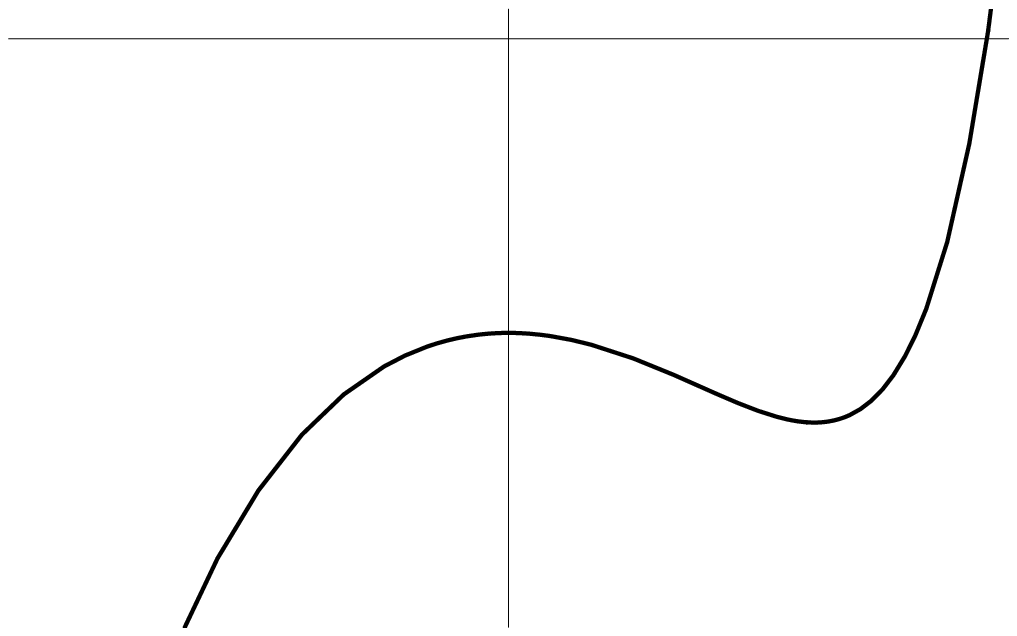} ~~~~
 \includegraphics[width=4.3cm]{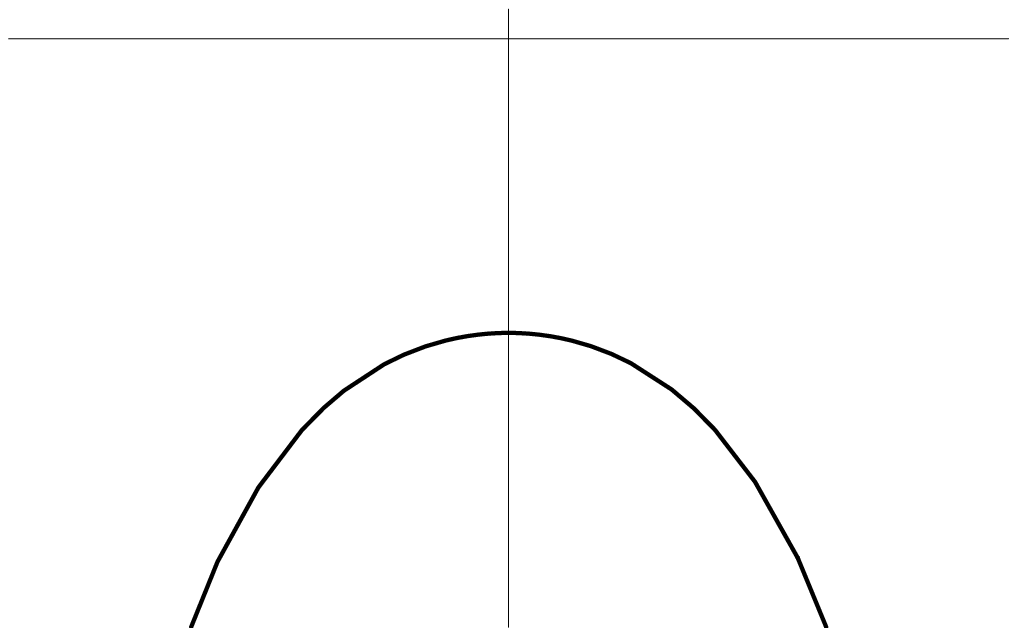} ~~~~
 \includegraphics[width=4.3cm]{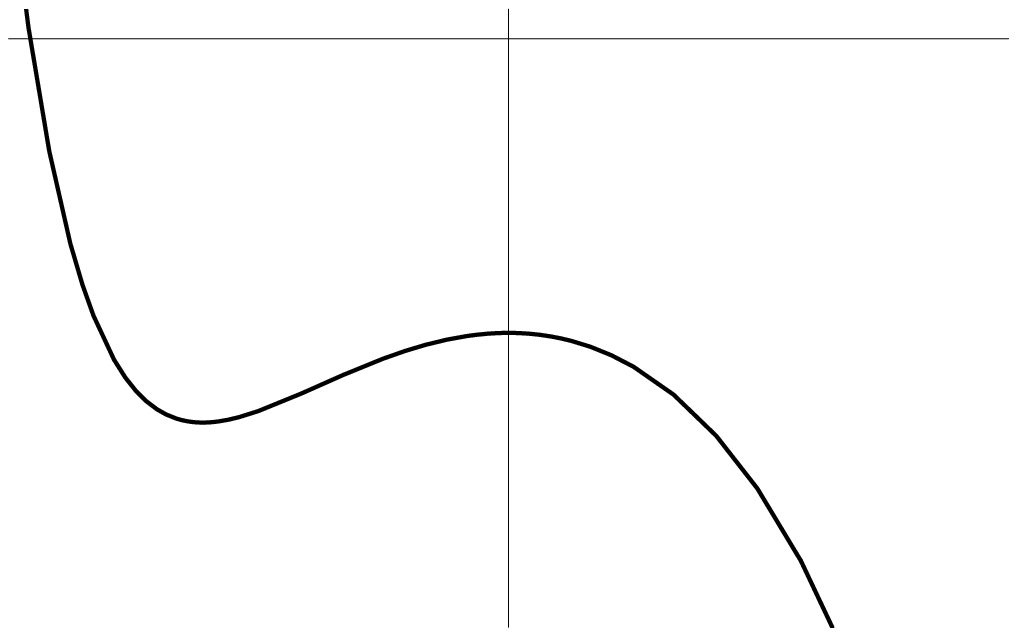}
}
\begin{picture}(0,0)(0,0)
\put(51,-8){\small $0<k<\frac{1}{\sqrt{3}}$}
\put(72,103){\footnotesize $V$}
\put(140,87){\scriptsize $\phi$}
\put(189,-8){\small $\frac{1}{\sqrt{3}} \le k \le \frac{2}{\sqrt{3}}$}
\put(217,103){\footnotesize $V$}
\put(288,87){\scriptsize $\phi$}
\put(335,-8){\small $\frac{2}{\sqrt{3}}<k<\infty$}
\put(364,103){\footnotesize $V$}
\put(433,87){\scriptsize $\phi$}
\end{picture}
\vspace{7mm}
\caption{\small The behavior of the potential $V$ for various
  values of $k$. For all $k$ the potential has a local maximum at
  $\phi = 0$ so that $V(0) = -6/L^2$. 
  Our solutions represent a ``flow'' from AdS at the top of the local
  maximum at $\phi=0$ towards $\phi \to \infty$ when $q>0$, and
  towards  $\phi \to -\infty$ when $q<0$.}    
\label{fig:V}
\end{figure}

When $1/\sqrt{3} \le k \le 2/\sqrt{3}$, the maximum at $\phi=0$ is global, but
for $0<k<1/\sqrt{3}$ or $k>2/\sqrt{3}$, the potential has a local
minimum located at
\bea
  \phi_\mathrm{min} 
  = \frac{3k}{2+3k^2} \log \left(\frac{3k^2-4}{2(3k^2-1)} \right) \, . 
\eea
Note that $\phi_\mathrm{min} < 0$ for $k>2/\sqrt{3}$, while 
$\phi_\mathrm{min} > 0$ for $0<k<1/\sqrt{3}$.
The behavior of the potential is sketched in figure \ref{fig:V}. The
local minimum appears to be of little significance for the solutions,
since the scalar is not stationary there due to the presence of a
non-vanishing electric field.

\subsubsection*{Solutions with $q>0$}
When $q>0$, the range of the coordinate $y$ is $0$ to $+$infinity: $y
\to+\infty$ is the asymptotic AdS region, and $y=0$ is the location of
a curvature singularity.
Since $H(y)$ in \reef{Hofy} is positive, the scalar $\phi$ \reef{phi1}
is non-negative. It flows from $\phi=0$ at the boundary
to $\phi \to +\infty$ at the singularity. 

It is possible to hide the curvature singularity behind an
event horizon for $q>0$ by turning on the non-extremality parameter $\mu$. The
horizon is located at the (largest) zero of the function $f$ in
\reef{fmu}. The conditions for the existence of a horizon are analyzed
in appendix \ref{app:horizon} and we summarize the result here:

\begin{itemize}
\item For $k>1/\sqrt{3}$ the solution has a single horizon whenever
  $\mu>0$. 
\item For $k=1/\sqrt{3}$ the existence of a horizon requires $\mu >
  q^2/L^2$. There is no inner horizon.
\item For $k<1/\sqrt{3}$ an event horizon requires $\mu \ge \mu_k(q,L)$,
  where $\mu_k(q,L)$ solves \reef{Htosolve}-\reef{H0plus}, as
  described in appendix \ref{app:horizon}. 
  Whenever $\mu > \mu_k(q,L)$ the solution has an inner horizon in
  addition to the event horizon. For $\mu = \mu_k(q,L)$ the horizons
  coincide, and the solutions are extremal but not fake BPS. 
\end{itemize}

Consider a solution with a horizon located at $y=y_h$. 
The Hawking temperature is 
\bea
  T_\rom{H} = \frac{1}{4\pi} \, \frac{f'(y_h)}{[H(y_h)]^{3p/2}}\, ,
\eea
and the entropy $S$, computed from the horizon area $A_\rom{H}$, is
\bea
  S = \frac{A_\rom{H}}{4G} 
    = \frac{\pi^2}{2 G} \, [H(y_h)]^{3p/2} \, y_h^3 \, .
\eea
We note in particular that the extremal non-fake-BPS solutions which
exist for $k<1/\sqrt{3}$ are characterized by having
$f(y_h)=f'(y_h)=0$. Hence these have zero temperature and finite
horizon area.

The superstar cases, $k = 1/\sqrt{3}$ and $2/\sqrt{3}$, are the
borderline cases for the behavior of the potential in figure
\ref{fig:V}. For $k = 2/\sqrt{3}$ the non-extremal superstars have
horizons when $\mu>0$. For $k = 1/\sqrt{3}$ a horizon exists when 
$\mu > q^2/L^2$.  


\subsubsection*{Solutions with $q<0$}
When $q<0$, the $y$-coordinate ranges from the boundary $y \to \infty$
to $y = \sqrt{|q|}$, where there is a naked singularity.
Note that $0<H(y)<1$, so $\phi$ is negative. Referring to figure
\ref{fig:V}, the scalar ``flows'' from AdS at the top of the local
maximum towards the singularity at $\phi \to -\infty$.

It is \emph{not} possible to hide the singularity behind a horizon for
any value of $\mu$ when $q<0$. Note that the non-extremality parameter
$\mu$ affects the electric field since 
$\tilde q = \sqrt{q(q+\mu)}$. A real electric field requires that
 \be \label{mucond}
 \tilde q^2 = |q| (|q|-\mu) > 0, \qquad {\rm i.e.} \qquad \mu <
 |q|
 \ee
However, with (\ref{mucond}), we see from (\ref{fmu}) that 
$f (y) > 0$ for $y \geq \sqrt{|q|}$.
Thus we conclude that for a physical electric field,
one cannot have a non-extremal solution in which the naked
singularity is shielded.

It was proposed in \cite{Gubser:2000nd} that a spacetime with a naked 
singularity may be considered physical if the solution generalizes
to one in which the singularity is hidden behind an event
horizon. This is not satisfied by the $q<0$
solutions, which also fail another criterium \cite{Maldacena:2000mw},
namely that $g_{tt}$ should not diverge at the singularity, since that
violates the UV/IR connection. Moreover, we show in section
\ref{s:mass} that the mass of the fake BPS solutions with $q<0$ is
negative. Each of these observations indicates that the solutions with 
$q<0$ are unphysical.


\setcounter{equation}{0}
\section{Mass and Charge from Holography}
\label{s:mass}

In this section we derive properties of the boundary field theory
from the AdS/CFT correspondence. We will use the formalism of
holographic renormalization in which field theory observables are
calculated from a properly renormalized on-shell action involving
the boundary limit of the bulk fields of our system. This
formalism was developed in several papers; for example see
\cite{hr1,hr2,hr3,hr4,hr6}. 

\subsection{Holographic stress-energy tensor}

The form of the boundary action depends
on the bulk fields and their mutual interaction. Fortunately the
relevant holographic observables were derived in \cite{holoren}
for the same bulk system we are studying, namely the metric, a
scalar dual to a $\D = 2$ operator, and massless gauge fields with
kinetic term modified by an exponential function of the scalar. In
fact for the specific values $k = 1/\sqrt{3}$ and
$2/\sqrt{3}$, the bosonic Lagrangian \reef{rs3}, with
\reef{qxyw}, agrees in full detail with the system studied in
\cite{holoren}. The
scalar sector of the matter system is invariant under the change
$k \to 2/(3k)$, but the gauge field sector differs for these two
values. Gauge field fluctuations were added to the system in
\cite{anatomy} and further studied in \cite{holoren}. There is an
$SO(4)\times SO(2)$ flavor symmetry, and it turns out that the
$SO(2)$ gauge field couples as in our system for the case
$k=2/\sqrt{3}$, while $SO(4)$ gauge fields correspond to
$k=1/\sqrt{3}$. 

In the holographic renormalization formalism of \cite{holoren}
the bulk metric is parameterized by
 \bea
  ds^2 = L^2 \frac{d\rho^2}{4\rho^2}
  + \frac{1}{\rho} g_{ij}(\r,x^i)\, dx^i dx^j \, ,
\eea
in which $\r$ is the radial variable, and the $x^i$ are an
arbitrary set of coordinates of the boundary at $\r =0$.
Thus our first task is to relate $\r$ to the radial
variable $y$ of \reef{newmet}. We need this relation near
the boundary, so we solve the equation
\bea
  \label{dydrho}
  \frac{dy}{d\rho} = - \frac{L}{2 \rho}\, f^{1/2} H^{-p/2} \, ,
\eea
as the series
\bea \lab{yrho}
  y \sim \frac{L}{\sqrt{\rho}} (1 + a_1 \rho + a_2 \rho^2 + \dots)
  \, ,
\eea
where the coefficients are given by
\bea \lab{yrho1}
  a_1 &=& -\frac{1}{4}-\frac{q}{(2+3k^2) L^2} \, ,\\
  a_2 &=& \frac{\mu}{8L^2}
          +\frac{q}{4(2+3k^2) L^2}
          -\frac{q^2(2-3k^2)}{4(2+3k^2)^2 L^4} \, .
\eea

After reexpression in terms of $\r$, the bulk fields
$g_{ij}(\r,x^i),~ A_\m(\r,x^i),~\phi(\r,x^i)$ of our solution have
expansions in the coordinate $\r$ which are determined by the boundary
limit of the field equations. In general, both powers of $\r$ and
ln$(\r)$ occur in these expansions, but it is obvious from the
solution \reef{introsoln} that there
are no logarithms in our case. We omit them in the expansions which
then take the simpler form:
\bea \lab{gexp}
g_{ij} &=& g_{ij}^{(0)} + g_{ij}^{(2)}\r + g_{ij}^{(4)}\r^2+\ldots\\
\lab{phiexp}
\phi &=& \phi^{0}\r + \phi^{(2)}\r^2 +\ldots\\
\lab{aexp}
A_i &=& A^{(0)}_i + A^{(2)}_i\r +\ldots
\eea
The leading term $g_{ij}^{(0)}$ of the expansion \reef{gexp} is the
spacetime metric for the boundary gauge theory. In our case this is
the metric
\be 
  \lab{bmet}
  ds_4^2 = g_{ij}^{(0)}dx^idx^j = -dt^2 + L^2 d\O_3 \, .
\ee
We have chosen the constant of integration in \reef{yrho}, so that
$L$ is the radius of the boundary $S^3$.
Note that the scalar "source rate" term, which would be
proportional to $\r \ln\r$ is absent for us and the leading term of
$\phi$ vanishes at the "vev rate".  For the gauge field, which is in
the gauge $A_\r=0$, the expansion of $A_i$ is more general than
actually occurs. Namely, only the component $A_t$ is non-vanishing,
and its source rate term $A^{(0)}_t$
 vanishes, leaving the vev rate $A^{(2)}_t \r$ as the leading term. 

We now apply the relevant formulas of Secs.~5-6 of \cite{holoren}
after conversion to our conventions.\footnote{The scalar $\phi$ of
  \cite{holoren} is multiplied by $1/\sqrt{2}$, and we reinstate
  dimensions by including a factor of $\frac{1}{4\pi G\, L}$.} Formula
(5.45) of \cite{holoren} gives the 1-point function of the field
theory stress tensor:   
\bea \lab{1pt}
  \langle T_{ij} \rangle
  &=& \frac{1}{4 \pi G \, L} \bigg\{ g^{(4)}_{ij}
   + \frac{1}{8} \Big[ \Tr (g^{(2)})^2
       - (\Tr g^{(2)})^2\Big]g^{(0)}_{ij}
   - \frac{1}{2} (g^{(2)})^2_{ij} \nonumber\\
  && \hspace{1.5cm}
   + \frac{1}{4} g^{(2)}_{ij} \Tr g^{(2)}
   + \frac{1}{6} (\phi^{(0)})^2 g^{(0)}_{ij}
   + \frac{3}{2} h^{(4)}_{ij} \bigg\}
  \, ,
\eea
in which all contractions are taken with $g_{ij}^{(0)}$. The
expression for $h_{ij}^{(4)}$ is given in (5.38) of \cite{holoren}.
It involves various contractions of the curvature tensor of the
boundary metric and vanishes for the metric \reef{bmet}. The effect
of the background gauge field was not considered in Sec.~5 of
\cite{holoren}, but it can be seen to vanish at the rate $\r^2$ at
the boundary and thus not contribute to the 1-point function
\reef{1pt}. Using the quite general holographic formula
$g_{ij}^{(2)} = - L^2 (R_{ij} - g^{(0)}_{ij} \, R/6)/2$, one can show
that the two terms in $\big[\ldots\big]$ cancel, so that \reef{1pt}
reduces to 
\bea  \lab{tij}
  \langle T_{ij} \rangle
  &=& \frac{1}{4 \pi G \, L} \bigg[ g^{(4)}_{ij}
   - \frac{1}{2} (g^{(2)})^2_{ij}
   + \frac{1}{4} g^{(2)}_{ij} \Tr g^{(2)}
   + \frac{1}{6} (\phi^{(0)})^2 g^{(0)}_{ij} \bigg]
  \, .
\eea
We can now evaluate this 1-point function by applying the
coordinate relation \reef{yrho} to the various contributions to
the bulk solution. 
We then 
obtain the stress tensor
\bea \lab{tijvev}
  \langle T_{tt} \rangle
  &=& \frac{1}{4 \pi G \, L} \bigg[ \frac{3}{16} + \frac{3 \mu}{4L^2}
   + \frac{3}{(2+3k^2) L^2}\,q \bigg] \, ,\\[2mm]
  \langle T_{\th\th} \rangle
  &=& \frac{1}{4 \pi G \, L} \bigg[ \frac{L^2}{16} + \frac{\mu}{4}
    + \frac{1}{(2+3k^2)}\,q \bigg] \, ,
\eea
for the field theory dual of the non-extremal solutions. 
The mass, 
$M = \int_{S^3} \langle T_{tt} \rangle 
= 2\pi^2 L^3\, \langle T_{tt} \rangle$, 
is then
\bea \lab{mass}
  M = \frac{\pi}{4G} \left[  \frac{3L^2}{8} + \frac{3 \mu}{2}
    + \frac{6}{2+3k^2}\,q \right] \, .
\eea
The first term proportional to $L^2$ is the Casimir energy
\cite{hr2}.
Substracting it we have
\bea
  M_0 = \frac{\pi}{4G} \left[ \frac{3 \mu}{2}
    + \frac{6}{2+3k^2}\,q \right] \, .
\eea
For extremal solutions one sets $\m=0$.
Note that using our asymptotic charge $q_\rom{elec}$ in \reef{EXTqelec},
the mass formula suggests a fake BPS bound
\bea
  M_0 \ge \frac{\pi}{4G} \, \sqrt{\frac{3}{2+3k^2}} \; q_\mathrm{elec} \, .
\eea
We will examine this using the Witten-Nester method in section
  \ref{s:WN}. 

Note that the trace of the stress-energy tensor vanishes exactly,
\be
  \< T^i_{~\,i} \> 
  = - \< T_{tt} \> + \frac{3}{L^2} \< T_{\th \th} \> = 0\, .
\ee 
This must be the case because the holograhic trace anomaly \cite{hr1}
is proportional to $R^{\mu \nu} R_{\mu \nu} - R^2/3$ and this vanishes
for the $\mathbb{R} \times S^3$ boundary metric.

Holographic renormalization also determines precise formulas for the
vevs of the operator $\co_\phi$ dual to the bulk scalar and the
conserved current $J^t$ dual to $A_t$. For the cases $k=1/\sqrt{3},
2/\sqrt{3}$, for which our solutions agree with the superstars,
$\co_{\phi}$ is a component of Tr($X^2)$ and $J^t$ is the time
component of a conserved $R$-current of $\cn=4$ SYM theory.
Formula (5.33) of \cite{holoren} gives the scalar vev
\be \lab{scvev}
  \<\co_{\phi}\> = \frac{1}{L^2} \, \sqrt{2}\, \phi^{(0)} 
  = \frac{3\sqrt{2}\,  k\, q}{(2+3k^2) L^4}  \, ,
\ee
where $\phi^{(0)}$ is given by \reef{phiexp}. 
From formula (6.77) of \cite{holoren} we find that
\be \lab{jvev}
\<J^t\> = \frac{2}{L^3} \, A^{(2)}_t = 
2 \sqrt{\frac{3}{2+3k^2}} \, \frac{\tilde{q}}{L^5} \, .
\ee
Note that 
\bea
  \frac{1}{2\pi^2}\int_{S^3} \<J^t\> = \frac{q_\rom{elec}}{L^2} \, ,
\eea
with $q_\rom{elec}$ given by \reef{elec}.


\subsection{Witten-Nester in fake supergravity}
\label{s:WN}

In this section we use the Witten-Nester method
to calculate the energy of our solutions. This method
determines the energy of a spacetime with respect to 
a background spacetime --- such as flat
Minkowski or anti-de Sitter space.

The Witten-Nester energy $E_\rom{WN}$ is defined as the asymptotic
boundary integral  
\bea
  \lab{WNbdr}
  E_\rom{WN} = \int_{\partial \Sigma}
    \star \hat{E} \, ,
  \hspace{1.5cm}
  i \hat{E}^{\m\n}&=& \bar{\e}_1 \G^{\m\n\r}\mathcal{D}_\r \e_2 -
  \bar{\e}_1\stackrel{\leftarrow}{\mathcal{D}}_\r  \G^{\m\n\r}\e_2 \, .
\eea
$\hat{E}$ is the Witten-Nester 2-form and $\mathcal{D}_\r\e$ is given
  by the fake gravitino transformation rule in \reef{Kspin}.
The spinors $\eps$ must asymptotically approach Killing spinors of the
  reference background, in this case AdS$_5$.  
Via Stokes' theorem, the boundary integral \reef{WNbdr} can be
converted to the bulk integral
\bea
  \lab{WNbulk1}
   E_\rom{WN} = \int_\Sigma d (\star \hat{E}) \, .
\eea
This results holds \emph{only if} there are no contributions from naked
  singularities or horizon boundaries. In our applications below we
  assume that $\mu$ is large enough that our solutions have regular
  horizons (cf.~discussion in section \ref{s:flow} and
  appendix \ref{app:horizon}). We further impose a condition on the Witten 
  spinor $\epsilon$ so that the contribution from the horizon boundary 
  vanishes. For details of this procedure, and discussion of the
  existence of Witten spinors, we refer to \cite{GHHP}.  

The standard approach of Witten-Nester is to show that the bulk
integral \reef{WNbulk1} is positive (semi)definite and vanishes only
for solutions which admit Killing spinors. 
The boundary integral
contains the conserved charges (mass, electric charge, angular
momentum etc). 
Combining the information from the bulk and boundary
integrals a BPS bound, or positive energy statement, is derived. 

A direct calculation of the bulk integral \reef{WNbulk1} with $\eps_1 =
\eps_2$ gives
\bea
  \lab{WNbulk2}
  \rom{Bulk:}
  \hspace{8mm}E_\mathrm{WN} &=&
  - i \int_\Sigma d\Sigma_\mu 
  \bigg[
  2\, \overline{\d \psi}_\nu \Gamma^{\m\n\r} \d \psi_\r 
  - \frac{1}{2}\, \overline{\d \lambda}\, \G^\m \, \d \lambda \\
  && \nonumber \hspace{2cm}
  - \frac{1}{2}  
  \, i\, \eps^{\mu \lambda\kappa\r\s} F_{\lambda\kappa}F_{\r\s}
  \Big( Y^2 - 48 X^2 + 72 b \, X Q^{-1} \Big)\,
   \bar{\eps} \, \eps 
  \bigg]\, .
\eea
This result depends on the form of the fake supergravity
  transformations, the identities \reef{qxyw} and the Einstein and
  gauge field equations. It is valid for any solution of the
  equations of motion. The first two contributions to the $F \wedge
  F$-term come from $\g$-matrix identities, while the last one comes from
  including a Chern-Simons term \reef{cs} with coefficient $b$.

Using $\bar{\eps} = \eps^\dagger i \ga^t$ and the Witten condition
$\G^a \mathcal{D}_a \eps = 0$, the first two terms of \reef{WNbulk2}
can be shown to be positive definite.  
Had it not been for the $F \wedge F$-term we would use this to derive
a general BPS bound relating mass and charge.  
The coefficient of the $F \wedge F$-term is identical to the first
condition of \reef{cs2} which was obtained by requiring the full
action to be linearly supersymmetric. As concluded in appendix
\ref{app:fake}, the condition can only be satisfied for 
$k=\pm 2/\sqrt{3}$ and $b=0$, or for $k=0$ and
$b=1/(6 \sqrt{6})$.\footnote{As discussed in section \ref{s:comp},
  the supersymmetric theory of the other superstar case,
  $k=\pm1/\sqrt{3}$, contains an extra gauge field which must be
  included in order to obtain a general Witten-Nester bound.}  

For the class of solutions with $F \wedge F = 0$, including
our static electrically charged generalized superstar solutions, we get a bound as follows.
The boundary integral can be evaluated by treating the spacetime of
interest as a fluctuation about the asymptotic background; in our case
this background is global AdS space. There are contributions from the
metric, the scalar and the gauge field. For our solutions the
Witten-Nester boundary integral gives 
\bea
  \lab{frombdr}
  \rom{Boundary:}
  \hspace{8mm}
  E_\mathrm{WN} = 
  2 \pi^2 \, v^\dagger_1 \left\{
  \bigg[ 
      \frac{3 \mu}{2}
       +\frac{6\, q}{2+3k^2} 
  \bigg] +  \frac{6\, \tilde{q}}{2+3k^2} \, i \ga^{t} \,
  \right\}v_2 \, ,
\eea
with $v_1$ and $v_2$ denoting unconstrained constant spinors. 
The first term of \reef{frombdr} is the mass.
The term with $i \g^t$ comes from the $F_{yt}$-term in the $S^3$
  components of $\mathcal{D}_\m \eps$ in \reef{Kspin} and is
  proportional to the charge.
The Witten-Nester argument tells us that the hermitean symmetric
quadratic form in \reef{frombdr} is non-negative, and we thus derive
the inequality
\bea \lab{bpswn} 
M_0 &\equiv& \frac{3 \mu}{2} + \frac{6\, q}{2+3k^2}
    ~\ge~ \frac{6}{2+3k^2} |\tilde{q}| = \sqrt{\frac{3}{2+3k^2}}
    |q_{\rm elec}|,
\eea
in which $q_{\rm elec}$ is the electric charge \reef{elec} of the
non-extremal solution. This is the bound anticipated from the
holographic calculation. 

One might have expected that the fake supergravity framework would
have allowed the derivation of a general bound relating energy and
charge. But this is false because the $F \wedge F$-terms in
\reef{WNbulk2} do not have the required positivity.
The same conclusion holds for any dimension $D\ge 5$, where a fifth
rank $\g$-matrix give the analogous non-positive 
$F_{\lambda\kappa}F_{\r\s}$-terms.  
In $D=4$, however, the fifth rank $\g$-matrix vanishes identically,
so linear supergravity is complete and a general BPS
bound can be derived. 


\setcounter{equation}{0}
\section{Discussion}
\label{s:disc}

We have extended the method of fake supergravity with the purpose of
exploring the AdS/CFT correspondence for field theories on $\mathbb{R}
\times S^3$. An abelian gauge field has been included in order to
obtain non-trivial solutions. The bulk
gauge field restricts the bulk Lagrangian and leaves only
a choice of a real constant $k$ which appears in the exponential
$e^{2k\phi}$ of the coupling between the scalar and the gauge field.
The fake supersymmetric electrically charged solutions of the
\mbox{$D=5$} first order 
flow equations, are generalizations of ``superstar'' solutions. For
special values of $k$ these are truly supersymmetric
superstars. Non-extremal generalizations include a Schwarzschild-like
parameter which makes it possible the hide the otherwise nakedly
singular source of the electric field behind an event horizon.

This work was originally motivated by the wish to find holographic
duals of renormalization group flows for field theories on 
$\mathbb{R} \times S^3$. As it 
turned out our solutions describe duals of states rather than
deformations of $\mathcal{N}=4$ SYM theory. It is possible that fake
supergravity will lead to new flows when carried out for a bulk theory
with more fields and perhaps with a solution ansatz which only
has the symmetry $\mathbb{R} \times S^3$ asymptotically.

Fake supergravity has been applied to describe
holographic renormalization group flows and to the problem of
(classical) stability. 
The applications also include a correspondence between domain walls
and cosmology solutions through analytic continuations
\cite{DWcosm}. Recently, a different formulation of fake supergravity
has been used to find first order flow equations for $D=4$
non-supersymmetric black hole solutions \cite{CD}. The diversity of
the applications demonstrate the power of the method.

Our work has illustrated the use of fake supergravity for
finding and solving first order flow equations, even in cases
where the action is linearly supersymmetric only for a certain class
of field configurations which include the solution ansatz. This in turn
revealed a limitation in the application of fake supersymmetry to
derivations of BPS bounds. The result indicates a connection
between obtaining general linear fake supersymmetry of the action and
achieving a BPS-type bound on the Witten-Nester energy. It would be
interesting to establish such a connection in more general contexts.

\subsubsection*{Acknowledgements}
We would like to thank Mirjam Cvetic, Michael Duff, Gary Gibbons, Matt
Headrick, Micheal Kiermaier, Albion Lawrence, Hong L\"u, Oleg Lunin, 
John McGreevy, Bob McNees, Ioannis Papadimitriou, Chris Pope, and Amit
Sever for useful discussions. 
HE is supported by a Pappalardo Fellowship in Physics at
MIT. DZF is supported by NSF grant PHY-0600465.
HL is supported in part by the Alfred~P.~Sloan
Foundation and U.S. Department of Energy (D.O.E) OJI grant.
In addition, all three authors are supported by the US Department of
Energy through cooperative research agreement DE-FG0205ER41360.


\appendix

\setcounter{equation}{0}
\section{Details of the fake supergravity
  construction}
\label{app:fake}
This appendix provides more information on the determination
of the various scalar functions in the fermion action and
transformation rules given in Section~\ref{s:fakegauge}.
Several $\g$-matrix identities are used in the calculations, such as  
\be \lab{gint}
  \g^{\m\n\r}
  \big(\g_\m\,^{\s\t} - 2(d-2) \d_\m^{~\s}\g^\t\big)F_{\s\t} 
  = - (d-1) \big( \g^{\n\r\s\t}F_{\s\t} + 2 F^{\n\r} \big) \, ,
\ee
in $D=d+1$ dimensions.

We begin by listing several of the $\bar{\lambda}(\cdots)\eps$ spinor
bilinears which appear in \mbox{$\d(S_b + S_f + S_\rom{gauge})$} and the
conditions that their vanishing imposes on the scalar functions:
\bea \lab{junk}
\begin{array}{rcrcl}
  \big(\bar{\l} \g^\m \g^{\r\s}F_{\r\s} D_\m\e\big)&~~~& 
  N&=&Y \, , \\[2mm]
  \big(\bar{\l} \g^{\m\r\s}F_{\r\s}\pa_\m \phi \, \e\big)&&
  P&=&(d-1)X-Y' \, , \\[2mm]
  \big(\bar{\l}\g_\m\pa_\n F^{\m\n}\e\big)&&
  \b \, Q &=& 2Y \, , \\[2mm]
  \big(\bar{\l} \g^\m F_{\m\n} \pa^\n\phi \, \e\big)&&
  \b\, Q' &=& 4Y' \, .
\end{array}
\eea
Eliminating $\b$ from the two relations in which it appears gives
\be \lab{juna}
\frac{Q'}{Q} = 2\frac{Y'}{Y}.
\ee
The analogues spinor bilinears involving the gravitino lead to the
additional conditions    
\bea 
\lab{junc}
\begin{array}{rcrcl}
  \big( \bar{\psi}_\m  \stackrel{\leftarrow}{D}_\nu
  (\g^{\m\n\r\s} F_{\r\s} + 2F^{\m\n}) \e\big) &&
  M&=&-2(d-1)X\, ,\\[2mm]
  \big(\bar{\psi}_\m \g^{\m\n\r\s} F_{\r\s}\pa_\n \phi \, \e\big)&&
  Y+N&=&4(d-1)X' \, ,\\[2mm]
  \big(\bar{\psi}_\m F^{\m\n}\pa_\n\phi \, \e\big)&~~~&
  \a\, Q' &=& -16(d-1)\,X' \, , \\[2mm]
  \big(\bar{\psi}_\m \pa_\n F^{\m\n} \e\big)&&  
  \a\, Q &=& -8(d-1)\,X\, .
\end{array}
\eea
The ratio of the two relations involving $\a$ gives
\be \lab{junb}
  \frac{Q'}{Q}=2\frac{X'}{X}.
\ee
Then, from \reef{juna} we learn that $X$ and $Y$ are
proportional. It is then convenient to impose
\be \lab{june}
  Y= 2(d-1) k \, X \, ,
\ee
with $k$ a constant.
{}From the two conditions involving $N$ and $Y$ above, we learn
that $Y = 2(d-1)X'$. Using \reef{june}, we find that
\be \lab{xep}
  X= c_1\, e^{k\,\phi}
\ee
in which $c_1$ is an integration constant which we will fix shortly. 

The vanishing condition for the coefficient of 
$(\bar{\psi}_\m \g^{\m\r\s}F_{\r\s}\e)$ is
\be \lab{seq}
  c + 4(d-1)(d-2)XW-2(d-1) YW'=0 \, .
\ee
Using \reef{june} for $Y$ and the exponential solution \reef{xep} for
$X$, this condition becomes a differential equation which determines
the  superpotential to be
\be \lab{supot1}
  W(\phi) = -\frac{c}{4 c_1 (d-1)\big( (d-2)+(d-1)k^2\big)} 
   \, e^{-k\phi} + c_2 \,e^{\frac{d-2}{(d-1)k}\phi}.
\ee

Next we discuss the several spinor bilinears of the form
$(\bar{\psi}\, \G F^2\e)$, in which $\G$ indicates a matrix of the
Clifford algebra of 5th, 3rd, or 1st rank. The 3 types are
independent, so their coefficients must vanish separately.
The 5th rank case 
is discussed below. The 3rd rank bilinear actually
vanishes due to index contractions. The 1st rank terms give us
the relation,
\be \lab{xyq}
 Q\,=\,4\big( 4(d-2)(d-1) X^2 + Y^2 \big)
 \,=\,16 \, c_1^2 \, (d-1) \big((d-2) +(d-1)k^2\big)\, e^{2k\,\phi},
\ee
after use of \reef{junk}, \reef{junc}, \reef{june} and \reef{xep}.

The functional form of all scalar functions in the ansatz
has been determined, and we now fix the integration
constants $c_1, c_2$ using some physical input. 
It is convenient to choose the constant $c_1$ so that
$Q(\phi) \to 1$ at the boundary. 
We then choose $c_2$ so that the 
stationary point of $W(\phi)$ occurs at $\phi=0$. 
We also normalize the superpotential so that the field equations
of the theory admit AdS$_D$ with scale $L$ as a solution.
With these conventions, the superpotential \reef{supot1} becomes 
\be \lab{supot}
  W(\phi) = \frac{1}{2L\big((d-2)+(d-1)k^2\big)}
  \Big[(d-2) e^{-k\, \phi} + (d-1) k^2
  e^{\frac{d-2}{(d-1)k}\phi} \Big] \, .
\ee
We regard $k$ and $L$ as the physical parameters of the model, and
express the U(1) coupling $c$ in terms of them. We can summarize these
relations among the parameters as
\bea \lab{params}
  c = -\frac{d-2}{2 L \sqrt{\frac{d-2}{d-1}+k^2}} \, ,
  ~~~~
  c_1 = \frac{1}{4 (d-1)\sqrt{\frac{d-2}{d-1}+k^2}} \, ,
  ~~~~
  c_2 = \frac{k^2}{2L \big(\frac{d-2}{d-1}+k^2\big)}\, .
\eea

With these choices, the potential takes the form
\bea
  V(\phi) = - \frac{d\, (d-1)}{2 L^2} - \frac{(d-2)}{L^2}\, \phi^2 
      + \dots \, 
\eea
when $\phi \to 0$, i.e.~near the AdS$_D$ boundary. This is the potential
of massive scalar in AdS$_D$ with mass $m^2 = -2(d-2)/L^2$. The BF
bound of AdS$_D$ is $m_\rom{BF}^2 = - d^2/(4L^2)$, so for $D=5$
we saturate the bound $m^2 = m_\rom{BF}^2$, whereas for $D=4$ or
$D>5$, the mass is strictly above the bound, $m^2 >
m_\rom{BF}^2$. Thus for all $D \ge 4$, fake supergravity has lead us
to theories with stable potentials 
$V$.\footnote{In an AdS/CFT context the bulk scalar generates a 
  deformation of the CFT by an operator of dimension $d-2$ such as a
  scalar mass deformation $M^2 \varphi^2$.} 

The vanishing conditions for 10 spinor bilinears have been used so
far to determine all scalar functions of the initial
ansatz. There are several other bilinears with at most rank 4
$\g$-matrices whose vanishing conditions can be seen to be satisfied
using previous results. These do not impose new conditions.

We do discuss briefly the issue of rank 5 $\g$-matrices which appear
in the calculation as $\G^{\m\n\l\r\s} F_{\m\n} F_{\l\r}$. For $D=4$
these terms of course vanish identically and no new conditions
appear. We treat the cases $D=5$ and $D>5$ separately.

For $D=5$, the 5th rank $\g$-matrix is proportional to the Levi-Civita
symbol, $\g^{\m\n\l\r\s} = i \epsilon^{\m\n\l\r\s}$, and so
these terms enter in the same form as supersymmetry variations of a
Chern-Simons term. Chern-Simons terms are usually present
in 5-dimensional supergravity theories. Thus we might expect it
necessary to add the bosonic term
\be \lab{cs}
  \cl_{CS} = b\, \e^{\kappa\m\n\r\s}A_\kappa F_{\m\n}F_{\r\s}
\ee 
to the Lagrangian of the fake supergravity model. Its supersymmetry
variation is 
\bea \lab{csvar}
  \d\cl_{CS} &=& - 3 i \, b\,   \e^{\kappa\m\n\r\s}
 \Big[ \a\, 
    \big(\bar{\e}\,\psi_\m - \bar{\psi}_\m \e\big)
    + \b \, \big(\bar{\e}\, \G_\m\l+\bar{\l}\, \G_\m\e\big)
    \Big]F_{\m\n}F_{\r\s}\, . 
\eea
In fact the supersymmetry variation $\d S_\rom{F}$ contains the spinor
bilinears such as $(\bar{\psi}_\t \g^{\t\m\n\r\s} F_{\m\n}F_{\r\s}\e)$ and
$(\bar{\l}\g^{\m\n\r\s}\e)$ which take the same form as the two
terms of \reef{csvar}  when the duality relations of the 5-dimensional
Clifford algebra are used. The coefficents of each bilinear are
quadratic in the scalar functions of the model. Assuming that
$\cl_{CS}$ is present, the cancellation conditions of the gravitino
and gaugino terms are
\bea 
  \lab{cs2}
  \hspace{-1cm}
  D=5: \hspace{1.5cm}
  \begin{array}{rcl}
  Y^2 - 48 X^2 -3 b\, \a &=&0 \, , \\[1mm]
  YY' -8XY -3 b\, \b &=&0 \, .
  \end{array}
\eea
The second condition is the derivative of the first if and only
if $b$ is constant, ie. independent of $\phi$. Of course, gauge
invariance requires constant $b$. However, using the last relation of
\reef{junc} together with \reef{june} and \reef{xep}, one finds that 
there are only two solutions to \reef{cs2}, namely 
(1) $k=\pm 2/\sqrt{3}$ and $b=0$, and (2) $k=0$ and
$b=1/(6\sqrt{6})$. Both these cases correspond to consistent
truncations of the $D=5$ supersymmetric $U(1)^3$ theory \cite{duff},
as discussed in section \ref{s:comp}. 

Thus we obtain a complete fake supergravity model only
for these cases. However, complete linear local supersymmetry is not
really required for application to solutions with purely electric
field, since the two spinor bilinears themselves vanish if $F_{rt}$  
is the only non-vanishing component of the field strength. The
fake supergravity approach can succeed even if the complete
linear local supersymmetry fails, provided that
$\d(S_b+S_f+S_\rom{gauge})$ vanishes for the class of solutions under
study. 

For $D>5$, the condition that all 5th rank $\G$-matrices cancel is
\bea
  \lab{5thrank}
  Y^2 = 4\, d\, (d-1) X^2  \, .
\eea
This selects the values 
\bea
  D>5: \hspace{1cm}  
  k = \pm \sqrt{\frac{d}{d-1}}
\eea
Again, for the purpose of fake supergravity, it is only necessary to
impose the condition \reef{5thrank} if $F \wedge F$ is non-vanishing
for the solutions of interest. 

\vspace{2mm}
It is somewhat surprising that the matrix structure of the
superpotential ${\bf W}(\phi)$, which is required for AdS$_d$
sliced domain walls does not appear in our study. In fact the ansatz
\reef{sga}, \reef{trf2} accommodates similar matrices at several
places. For example, the spinors could be doubled and $X$ and $Y$
replaced by matrices. However, fake supergravity is modeled on
real $D=5$, $\cn=2$ supergravity in which spinor doubling occurs because
real $D=5$ supergravity requires symplectic Majorana spinors. So we
consulted the form of the fermion transformation rules in
\cite{dutch} and found that analogues of $X$, $Y$ are diagonal in
the symplectic indices. Thus we assumed that $X$, $Y$ are scalars,
rather than matrices. Then, compatibility of \reef{seq} with the
matrix constraint \reef{com} tells us that $W(\phi)$ is also
scalar.


\setcounter{equation}{0}
\section{Conditions for existence of a horizon}
\label{app:horizon}

Our fake BPS solutions are all nakedly singular. The non-extremal
solutions have horizons when the function $f$ has a zero for $y>0$. It
is useful to examine the condition $f(y)=0$ using $H = 1 +
q/y^2$ as a variable instead of $y$, and to formulate the problem in
terms of the function 
\bea
  g(H) 
  ~=~ \frac{q^2}{L^2 \, y^2} f(y) \Big|_{y \to H}
  ~=~ \bar{q}^2 H^{3p} +\bar{q} (H-1)  - \bar{\mu} (H-1)^2 \, ,
\eea
where we have introduced dimensionless parameters
$\bar{q} = q L^{-2}$ and $\bar{\mu}=\mu L^{-2}$.
We focus on the case of $\bar{q} >0$ and  $\bar{\mu}>0$.
The condition for having a horizon is then that there exists an $H_0 >
1$ such that  
$g(H_0) = 0$. Note that $g(1)=\bar{q}^2 > 0$. Depending on the
behavior of $g$ for $H \to \infty$ there three cases: 
\begin{enumerate}
  \item{Case $k > \frac{1}{\sqrt{3}}$  (i.e.~$0<p<2/3$):}
  For $H \gg 1$, we have $g(H) \sim - \bar{\mu} H^2 < 0$. In this case
  $g$ always has a single zero for $H>1$, and so for any values of $q
  > 0$ and $\mu > 0$, the solution has a horizon. 
  \item{Case  $k =\frac{1}{\sqrt{3}}$ (i.e.~$p=2/3$):}
  For $H \gg 1$, we have 
  $g(H) \sim (\bar{q}^2 - \bar{\mu}) H^2 + (\bar{q}+ 2 \bar{\mu}) H +
  \dots$. 
  So if $\bar{\mu} > \bar{q}^2$, $g$ has a zero. 
  When $\bar{\mu} \le \bar{q}^2$, it is straightforward to see that
  $g$ has no local extrema for $H > 1$, and therefore $g>0$ for $H >
  1$. In conclusion, for $k =\frac{1}{\sqrt{3}}$ the solutions 
  have horizons only if  $\bar{\mu} > \bar{q}^2$, i.e.~if 
  $\mu > q^2/L^2$.
\item{Case $0<k < \frac{1}{\sqrt{3}}$  (i.e.~$2/3<p<1$):} 
  Since $g(H) \to \bar{q}^2 H^{3p} > 0$
  for $H \gg 1$, the existence of a zero of $g$ requires that $g$
  has a local minimum $H_\mathrm{min}>1$ such that $g(H_\mathrm{min})
  \le 0$. Note that $g'(1) = \bar{q} (1 + 3 p \bar{q}) > 0$, so $g$
  can only have a local minimum if it also has a local maximum. So we
  need $g'$ to have two separate zeroes. That in turn
  requires that $g''$ has a zero for $H>1$. Solving $g''=0$ gives
\bea
  \label{gpp}
  H^{3p-2} = \frac{2 \bar{\mu}}{3p (3p-1) \bar{q}^2} \, .
\eea
Requiring $H^{3p-2}>1$ in \reef{gpp} gives 
\bea
  \label{cond1}
  \bar{\mu} > \frac{3}{2}p (3p-1) \bar{q}^2 \, 
\eea
as a necessary condition for having a horizon.
Note that \reef{cond1} implies $\bar{\mu} > \bar{q}^2$.

The condition is \reef{cond1} not sufficient, so we push the analysis
further to solve the ``extremal'' case where $g$ has a minimum at
$g=0$; i.e.~we solve the system $g(H)=0$ and $g'(H)=0$.

Zeroes of $g'$ are solutions $H > 1$ to the equation
\bea
  \label{H0def}
  H^{3 p-1} = \frac{2 (H-1) \bar{\mu} - \bar{q}}{3 p \bar{q}^2} \, .
\eea
Plugging this into $g$ gives 
\bea
  \label{gH0}
  3 p\, g(H) = - (3p -2) \bar{\mu} H^2
   + (3p-1) (\bar{q} + 2 \bar{\mu}) H 
   - 3 p (\bar{q} + \bar{\mu}) \, .
\eea
Let the two zeroes of \reef{gH0} be $H_0^{\pm}$, with $H_0^- < H_0^+$. One
   finds that $H_0^\pm$ are real and satisfy $H_0^\pm > 1$. However,
   for $H=H_0^-$, the RHS of \reef{H0def} is negative, so we discard
   this as a solution. Setting $H=H_0^+$ in \reef{H0def} gives 
   an equation that can be used to determine $\bar{\mu}$ numerically
   for given $\bar{q}$ and $p$:
\bea
  \label{Htosolve}
  (H_0^+)^{3 p-1} = \frac{2 (H_0^+-1) \bar{\mu} - \bar{q}}{3 p \bar{q}^2} \, ,
\eea
with 
\bea
  \label{H0plus}
  H_0^+ = \frac{(3p-1)(\bar{q}+2\bar{\mu}) 
  + \sqrt{(3p-1)^2 \bar{q}^2 + 4 \bar{\mu} ( \bar{q} + \bar{\mu})}}
  {2(3p-2) \bar{\mu}} \, .
\eea
It can be shown that whenever the
condition \reef{cond1} holds, the RHS of \reef{Htosolve} is greater
than 1. 

To summarize, for given $\bar{q}$ and $0<k< \frac{1}{\sqrt{3}}$, $\mu$
needs to be 
sufficiently large in order for a horizon to exist. Condition
\reef{cond1} is a necessary, but not sufficient, condition. Solving
\reef{Htosolve} and \reef{H0plus} numerically for given $\bar{q}$ and
$p$ gives the value $\bar{\mu}_k(\bar{q})$ for the ``extremal'' black
hole solution which has minimum energy above extremality for given
charge $\bar{q}$. Re-introducing the AdS scale $L$, we denote the
above $\mu$-bound by $\mu_k(q,L)$. 
When $\mu > \mu_k(q,L)$, the function $g$ has two zeroes and the
solutions have both an inner and an outer horizon. 

As an example, consider $k=\sqrt{2}/3$ (i.e.~$p=3/4$) and  $\bar{q}=1$. 
Then condition  
\reef{cond1} gives $\bar{\mu} > 45/32 \approx 1.4$, whereas solving
\reef{Htosolve} shows that the extremality bound is
$\bar{\mu}_{k=\sqrt{2}/3}(\bar{q}=1) \approx 2.3$ for this example. 

\end{enumerate}


\section{Fake Killing spinors}
\label{s:fakeKS}

We consider first Killing spinors for pure AdS$_5$ in the coordinates
obtained as the $q=\mu=0$ limit of our solutions.
Their form
serves as a useful starting point for the construction of fake
Killing spinors for the extremal solutions above, and they 
play a direct role in the Witten-Nester energy computation.

In the diagonal frame for the AdS$_5$ metric \reef{adsy}, it is
straightforward to show that
the AdS$_5$ Killing spinor $\eps_0$ can be written
\bea 
  \lab{ksads}
   \eps_0 =
   \Big[ g_+(y) P_+ + g_-(y) P_- \Big]\, e^{-\frac{1}{2L}\ga^t\, t}\,  v \, ,
\eea
where we have introduced the projectors $P_\pm = \frac{1}{2} (1\pm \ga^y)$
and functions
\bea
  \lab{gpm}
  g_\pm(y)
  =  \left[ \sqrt{f_0(y)} \mp \sqrt{f_0(y)-1}
  \right]^{1/2} ,~~~&&~~~f_0(y)= 1+\frac{y^2}{L^2}.
\eea
The spinor $v$ is one variant of the $S^3$ Killing spinors obtained
in \cite{pope}. It satisfies
\bea
  \lab{vspinor}
  \bar{\nabla}_i v = - \frac{1}{2} \bar{\ga}_i \ga^y v \, ,
\eea
where $\bar{\nabla}_i$ and $\bar{\ga}_i$ denote the derivatives
(including spin-connections) and $\ga$-matrices for the unit
$S^3$. 

The two presentations of the AdS$_5$ metric \reef{adsr} and
\reef{adsy} differ by the relation $y=L \sinh(r/L)$ of their
radial coordinates. In the coordinate $r$, the Killing spinor
\reef{ksads} can be rewritten as the simple expression
\be \lab{ks2}
\e_0 \,=\, e^{\frac{1}{2L}\g^5 r} \, e^{-\frac{1}{2L}\ga^t\,
  t}\,  v \,,
\ee
which can be shown to agree with the form in Appendix E of \cite{cheng}.

The extremal solutions specified by \reef{newmet}-\reef{newat} admit fake Killing spinors. To find them, we make the ansatz
\bea
 \eps =  \Big[ f_+(y) P_+ + f_-(y) P_- \Big]\, u(t,\th_i) \, ,
\eea
where $\th_i$ are the coordinates of the sphere $S^3$.
This ansatz must satisfy the fake Killing spinor equations
\reef{Kspin} (using \reef{kdil}-\reef{kth}) when the solution data
\reef{newmet}-\reef{newat} is inserted. 
Analyzing the near-AdS limit of the equations, we find that the condition 
\bea
  \lab{halfsusy1}
  i \ga^t u = u \, .
\eea
is required. We expect that our solutions are at most half fake BPS,
and we therefore 
impose the condition \reef{halfsusy1} when constructing exact fake
Killing spinors.   
With this, it can be shown that the equations
$\hat{\mathcal{D}} \eps=0$ and $\mathcal{D}_y \eps=0$ are
solved if
\bea
  \lab{fpm}
  f_\pm(y) =  H(y)^{-\frac{1}{2+3k^2}}
  \left[ \sqrt{f(y)} \mp \sqrt{f(y)-1} \right]^{1/2} \, ,
\eea
where $f(y)$ is given in \reef{newf}.

Next $\mathcal{D}_t \eps=0$ reduces to
\bea
  \lab{dteqn}
  i\, \pa_t u = \frac{1}{2L} u \, ,
\eea
with solution
\bea
  u(t,\th^i) = e^{-i\frac{1}{2L} t} \, v(\th^i) \, .
\eea
The spinor $v$ depends only on the $S^3$ coordinates.
Finally, $\mathcal{D}_i \eps=0$, with $i$ running over the $S^3$
coordinates $\th_i$, simplifies to the conditions
\bea
 P_\pm ( \bar{\nabla}_i v \mp \frac{1}{2}\bar{\ga}_i v) &=& 0 \, ,
\eea
which are simply equivalent to $S^3$ Killing spinor equations \reef{vspinor}.

Thus our solutions admit fake Killing spinors
\bea \lab{fks}
   \eps = \Big[ f_+(y) P_+ + f_-(y) P_- \Big]\, e^{-i\frac{1}{2L} t}\,  v \, ,
\eea
with $f_\pm$ given by \reef{fpm} and with $v$, satisfying
\reef{vspinor}, a Killing spinor on the unit $S^3$ \cite{pope},
constrained by the half fake BPS projection condition  
\bea \lab{proj}
  \label{halfsusy2}
  i \ga^t v = v \, .
\eea
Note that $f_\pm = g_\pm$ for $q = 0$, with  $g_\pm$ in
\reef{gpm}; in particular our fake Killing spinors asymptotically
approach the AdS Killing spinors \reef{ksads}.

The Killing spinor bilinears $(\bar{\e}\,\G^\m\e)$ are Killing
vectors of the bulk metric \reef{newmet}, whose isometry group
is $\mathbb{R}\times SO(4)$. This is the compact subgroup of the
$SO(4,2)$ group 
whose Killing vectors are denoted by $K_{[AB]}^\m$ and whose
spinor representation has generators $\g^{[AB]}$ given by
\bea \lab{spinrep}
\g^{[ab]} &=& \half \g^a\g^b
~~~~~~a,b=0,1,2,3,4~~~~~~{\rm rotations~and~boosts}\\
\g^{[a5]} &=& \half \g^a ~~~~~~~{\mathrm{energy~and~``momentum''}}.
\eea The spinor bilinears for both pure AdS$_5$ and $\mathbb{R} \times S^3$
solutions are then given by \bea \lab{spinbil}
(\bar{\e}_0\G^\m\e_0) &=& \bar{v}\g^{[a5]}v\,K_{[a5]}^\m +\half
 \bar{v}\g^{[ab]}v\,K_{[ab]}^\m\\
(\bar{\e}\G^\m\e) &=& \bar{v}\g^{[05]}v\,K_{[05]}^\m +\half
 \bar{v}\g^{[ab]}v\,K_{[ab]}^\m, ~~~~ a,b = 1,2,3,4.
\eea Note the restriction to energy and spatial rotations in the
$\mathbb{R}\times S^3$ case which is due to the projection constraint
\reef{halfsusy2}.


\end{document}